\documentclass[10pt,floats,epsfig]{iopart}  
\usepackage{enumitem}
\usepackage{graphicx}
\usepackage{amssymb}
\usepackage{bbm}
\usepackage{color}
\def\be{\begin{equation}}
\def\ee{\end{equation}}
\def\bea{\begin{eqnarray}}
\def\eea{\end{eqnarray}}
\begin{document}

\def\tcr{\textcolor{red}}

\title {The dynamical viability of scalar-tensor gravity theories}

\author{Nishant Agarwal and Rachel Bean\\
Department of Astronomy, Cornell University, Ithaca, NY 14853, USA\\
}
\begin{abstract}
We establish the dynamical attractor behavior in scalar-tensor theories of dark energy,  providing a powerful framework to analyze classes of theories, predicting common evolutionary characteristics that can be compared against cosmological constraints. In the Jordan frame the theories are viewed as a coupling between a scalar field, $\Phi$, and the Ricci scalar, $R$, $F(\Phi)R$. The Jordan frame evolution is described in terms of dynamical variables $m \equiv d\ln F/d\ln \Phi$ and $r \equiv -\Phi F/f$, where $F(\Phi) = d f(\Phi)/d\Phi$.  The evolution can be alternatively viewed in the Einstein frame  as a general coupling between scalar dark energy and matter, $\beta$. We present a complete, consistent picture of evolution in the Einstein and Jordan frames and consider the conditions on the form of the coupling $F$ and $\beta$ required to give the observed cold dark matter (CDM) dominated era that transitions into a late time accelerative phase, including transitory accelerative eras that have not previously been investigated. We find five classes of evolutionary behavior of which four are qualitatively similar to those for $f(R)$ theories (which have $\beta=1/2$). The fifth class exists only for $ |\beta| < \sqrt{3}/4$, \textit{i.e.} not for $f(R)$ theories. In models giving transitory late time acceleration, we find a viable accelerative region of the $(r,m)$ plane accessible to scalar-tensor theories with any coupling, $\beta$ (at least in the range $|\beta| \leq 1/2$, which we study in detail), and an additional region open only to theories with $|\beta| < \sqrt{3}/4$.
\end{abstract}
\maketitle

\section{Introduction}
\label{One}

The current accelerated expansion of the universe has led to a variety of new theories in cosmology under a common descriptor of {\it dark energy}. The acceleration was first observed using data from Supernovae Ia \cite{Perlmutter:1998np,Riess:1998cb,Riess:2004nr,Astier:2005qq,Riess:2006fw} and has been confirmed by  complementary observations including WMAP data on Cosmic Microwave Background anisotropies (CMB) \cite{Spergel:2006hy,Jarosik:2006ib,Hinshaw:2006ia,Page:2006hz}, data from the SDSS on Large Scale Structure formation (LSS) \cite{Tegmark:2003ud, Sanchez:2005pi,Tegmark:2006az}, baryon acoustic oscillations \cite{Eisenstein:2005su,Percival:2007yw} and weak lensing \cite{Jain:2003tba, Semboloni:2005ct,Kitching:2006mq}. 

The simplest way of introducing an acceleration into Einstein's equations is to add a cosmological constant so that the decelerating expansion (which we naturally expect because of the attractive force of gravity) gets nullified and over-compensated for. The cosmological constant though suffers from the problem of fine tuning since its energy density is $10^{120}$ times smaller than what we might expect if it related to Planck scale phenomena. Other models of dark energy include introducing a scalar field (such as minimally coupled quintessence models \cite{Wetterich:1987fm,Ratra:1987rm,Fujii:1989qk,Chiba:1997ej,Carroll:1998zi}, coupled quintessence models, \cite{Uzan:1999ch,Amendola:1999er,Bean:2000zm,Bean:2001ys,Copeland:2006wr}, K-essence \cite{Chiba:1999ka,ArmendarizPicon:2000dh,ArmendarizPicon:2000ah}, Chameleon fields \cite{Khoury:2003aq,Khoury:2003rn,Brax:2004qh} etc), Phantom dark energy \cite{Starobinsky:1999yw,Caldwell:1999ew,Caldwell:2003vq}, Chaplygin gases \cite{Kamenshchik:2001cp,Bilic:2001cg,Bento:2002ps,Sandvik:2002jz,Bean:2003ae}, topological defects \cite{Battye:1999eq}, and many others. A recent review on dark energy can be found in Ref. \cite{Copeland:2006wr}. 

An alternate approach to explaining the observed acceleration, without recourse to a new form of energy, is to consider large scale modifications to gravity coming from scalar-tensor gravity \cite{citeulike:2490641, Carroll:2003wy, Chiba:2003ir, Dolgov:2003px,Nojiri:2003rz,Allemandi:2004wn,Capozziello:2005ku, Carloni:2004kp,Amarzguioui:2005zq, Nojiri:2006ri, Amendola:2006kh, Capozziello:2006dj,Amendola:2006eh, Brookfield:2006mq, Bean:2006up,Capozziello:2007iu} or higher order gravity theories  \cite{Carroll:2004de, Nojiri:2005jg,Nojiri:2005am,Amendola:2005cr,Cognola:2006eg,Cognola:2006sp,Cognola:2007vq,Li:2007jm} such as modified Gauss-Bonnet gravity. One of the earliest modifications \cite{Carroll:2003wy} added a $1/R$ term to the gravitational action. Such modifications to gravity seem to conflict with local solar system tests of gravity, however general $f(R)$ gravity models remain interesting because there are ways of evading local gravity constraints \cite{Khoury:2003aq,Khoury:2003rn,Brax:2004qh,Hu:2007nk}. 

Dynamical attractors in dark energy theories provide a powerful way to analyze classes of theories. They  predict common evolutionary characteristics, with minimal dependence on fine-tuning of initial conditions \cite{Copeland:1997et,Uzan:1999ch, Amendola:1999er,Copeland:2006wr}, with which we can critique theories in light of cosmological constraints on the background evolution \cite{Amendola:2006kh, Amendola:2006we} and also the growth of perturbations and implications for large scale structure \cite{Bean:2006up}. 

In this paper we apply dynamical attractor analysis to general scalar-tensor theories.  In the Einstein frame theories are described by a coupling between scalar dark energy and matter, parameterized by $\beta$, and in the Jordan frame by a general coupling between a scalar field and the Ricci scalar,  $F(\Phi)R$; $\beta=0$ and $F(\Phi)\equiv 1$ for General Relativity. We have structured our paper in the following way. Section \ref{Two} contains the equations for background evolution in the Einstein frame  (Sec. \ref{Twoa}) and Jordan frame (Sec. \ref{Twob}) for generally coupled models. We discuss the two alternative perspectives given by the two frames (Sec. \ref{Twoc}) and the form of the Jordan frame potential we consider (Sec. \ref{Twod}). In section \ref{Three} we set up all of the equations that we need to study the critical points of dynamical evolution in the Jordan frame (Sec. \ref{Threea}) and Einstein frame (Sec. \ref{Threeb}), discuss the mapping of critical points from one frame to the other (Sec. \ref{Threec}) and how stability is assessed (Sec. \ref{Threed}). In section \ref{Four} we present the dynamical attractor solutions (Sec. \ref{Foura}) and compare our solutions with those previously obtained in the literature for $f(R)$ and non-minimally coupled quintessence theories (Sec. \ref{Fourb}). We perform a detailed analysis of the stability of all critical points (Sec. \ref{Fourc}) and classify scalar-tensor theories into different classes (Sec. \ref{Fourd}). We then discuss behavior if we relax the condition of a stable accelerated universe to include saddle critical points (Sec. \ref{Foure}). Finally we summarize and conclude in section \ref{Five}.
 
\section{Cosmic evolution}
\label{Two}
\subsection{Evolution in the Einstein frame}
\label{Twoa}

We start with the action in the Einstein frame (EF),
\begin{equation}
	\fl S = {\int} d^{4}x \sqrt{-\tilde{g}} 
	\left[
		\frac{1}{2\kappa^{2}}\tilde{R} - \frac{1}{2} (\tilde{\nabla} \phi)^{2} - \tilde{V}(\phi)
	\right] + {\int} d^{4}x \sqrt{-\tilde{g}} \tilde{\mathcal{L}}_{m}\left(e^{-2\sqrt{2/3} \kappa \beta \phi} \tilde{g}_{\mu\nu}\right)\label{EF}
\end{equation}
where $\kappa^{2} \equiv 8\pi G$, $\phi$ is a scalar field which is non-minimally coupled to matter through a coupling strength $\beta$. We denote all quantities in the EF with a tilde, such as the metric, $\tilde{g}_{\mu\nu}$, Ricci scalar, $\tilde{R}$, scalar potential, $\tilde{V}(\phi)$ and the matter Lagrangian, $\tilde{\cal L}_{m}$. The action in the EF is a standard General Relativity action with a non-minimal coupling between the scalar field and matter. For a Robertson Walker metric with flat geometry described by a cosmic expansion factor, $\tilde{a}$, evolving in physical time, $\tilde{t}$, the Friedmann, acceleration, scalar field and matter fluid equations are respectively,
\begin{eqnarray}
	&& \tilde{H}^{2} = \frac{\kappa^{2}}{3} \left( \tilde{\rho}_{m} + \frac{1}{2}\left(\frac{d\phi}{d\tilde{t}}\right)^{2} + \tilde{V} \right), 
	\label{FriedmannEF} \\
	&& \frac{d\tilde{H}}{d\tilde{t}} = -\frac{ \kappa^{2}}{2} \left(\tilde{\rho}_{m} + \tilde{p}_{m} +\left(\frac{d\phi}{d\tilde{t}}\right) ^{2}\right), \\
	&& \frac{d^2\phi}{d\tilde{t}^2} + 3\tilde{H}\frac{d\phi}{d\tilde{t}} + \tilde{V}_{,\phi} = \sqrt{\frac{2}{3}} \kappa \beta (\tilde{\rho}_{m} - 3\tilde{p}_{m}), \label{field}\\
	&& \frac{d\tilde{\rho}_{m}}{d\tilde{t}} + 3\tilde{H} (\tilde{\rho}_{m} + \tilde{p}_{m}) = - \sqrt{\frac{2}{3}} \kappa \beta \frac{d\phi}{d\tilde{t}} (\tilde{\rho}_{m} - 3\tilde{p}_{m}).  
	\label{bkg}
\end{eqnarray}
where $\tilde{\rho}_{m}$ and $\tilde{p}_{m}$ are the EF matter density and pressure, $\tilde{H}\equiv d\ln \tilde{a}/d\tilde{t}$ and $\tilde{V},_{\phi}\equiv d\tilde{V}/d\phi$.  In this paper we are interested in the nature of cosmic evolution in the matter dominated and dark energy dominated eras. We therefore restrict our attention to pressureless matter, with $\tilde{p}_{m}=0$.  

Such a parameterization, using $\beta$ to describe a non-minimal coupling between matter and scalar, has been widely discussed in the literature, for example \cite{Uzan:1999ch,Amendola:1999er,Bean:2000zm,Bean:2001ys, Amendola:2006kh,Amendola:2006we,Copeland:2006wr,Bean:2007nx, Bean:2007ny}. This can include low temperature neutrinos which behave as nonrelativistic matter. The scalar coupling of the form shown in (\ref{bkg}) can hence give rise to mass varying neutrinos (`MaVaN's)  \cite{Fardon:2003eh,Afshordi:2005ym,Bjaelde:2007ki, Bean:2007ny}.

\subsection{Evolution in the Jordan frame}
\label{Twob}

We can re-express the action (\ref{EF}) in the Jordan frame (JF) as a  non-minimal coupling of the scalar  to gravity and minimally coupled matter, through a conformal transformation, redefining the metric, %
\begin{equation}
	{g}_{\mu\nu} \equiv e^{-2\sqrt{2/3} \kappa \beta \phi} \tilde{g}_{\mu\nu}.
\end{equation}
This leads to  a general Brans-Dicke type action for a scalar-tensor theory with a non-canonical kinetic term,
\begin{eqnarray}
	\fl S = {\int} d^{4}x \sqrt{-g} 
	\left[ 
		\frac{1}{2\kappa^{2}} F(\Phi)R - \frac{3(1-4\beta^{2})}{16\kappa^{2}\beta^{2}}\frac{1}{F(\Phi)} \left(\frac{\textrm{d}F}{\textrm{d}\Phi}\right)^{2} (\nabla \Phi)^{2} - V(\Phi) 
	\right] \nonumber \\
	+ {\int} d^{4}x \sqrt{-g} \mathcal{L}_{m}(g_{\mu \nu}) 
	\label{JFaction}
\end{eqnarray}
where we have introduced the function $F(\Phi)$ of scalar $\Phi$, which is wholly specified by $\phi$ and $\beta$,
\begin{equation}
	F(\Phi)\equiv e^{2\sqrt{2/3} \kappa \beta \phi}
	\label{redefinition}
\end{equation}
We require $F(\Phi)>0$ for the conformal transformation to be well defined. $\Phi(\phi)$ can be specified by choosing an explicit form for $F(\Phi)$ and inverting (\ref{redefinition}).
 
Pressure, density, time, and expansion factor in the two frames are related via $F(\Phi)$,
\begin{eqnarray}
		\fl p_{m} = F(\Phi)^{2}\tilde{p}_m, \; \rho_{m} =  F(\Phi)^{2} \tilde{\rho}_m, \; \textrm{d}t =  F(\Phi)^{-1/2} \textrm{d}\tilde{t}, \; a = F (\Phi)^{-1/2} \tilde{a}, \ \ \ \ \ \
	\label{EFJF}
\end{eqnarray}
with the potential in the JF related to the EF potential  by,
\begin{equation}
	V(\Phi) =  F(\Phi)^{2} \tilde{V}(\phi),
\end{equation}
where again, an explicit form for $V(\Phi)$ can be found given a specific choice of $F(\Phi)$ and hence $\Phi(\phi)$.

The Friedmann, acceleration, scalar field and energy conservation equations for the action in the JF are:

\begin{eqnarray}
	&& H^{2} =\frac{\kappa^{2}}{3F}\left(\rho_{m} + V(\Phi)\right)- \frac{H\dot{F}}{F}  + \frac{ (1-4\beta^{2}) }{16\beta^{2}}\frac{\dot{F}^{2}}{F^2} , \label{FriedmannJF} \\
	&& \dot{H} = -\frac{1}{2F} \kappa^{2}\rho_{m} -\frac{1}{2}\frac{ \ddot{F}}{F} + \frac{1}{2}\frac{H\dot{F}}{F} - 3\frac{(1-4\beta^{2})}{16\beta^{2}}  \frac{\dot{F}^{2}}{F^2}, \label{JFflu1}\\
	&& \ddot{F} + 3H\dot{F} = \frac{4\kappa^{2}\beta^{2}}{3} \left(\rho_{m} + 4V(\Phi) -2F \frac{V_{,\Phi}}{F_{,\Phi}}\right),\label{JFflu2}  \\
	&& \dot{\rho}_{m} + 3H\rho_{m} = 0. \label{JFflu3}
\end{eqnarray}
where dots denote derivatives with respect to physical time $t$ in the JF.
%

\subsection{The Einstein frame vs. Jordan frame perspectives}
\label{Twoc}

If a theory can be expressed in a frame in which all matter components are minimally coupled, such as in (\ref{JFaction}), then a conformal transformation to the Einstein frame will always result in the matter-scalar coupling being the same strength, $\beta$, for all types of matter. Some debate persists in the literature, in the case of such identical couplings, on whether EF or JF is the `physical' frame, see for example Refs. \cite{Faraoni:1999hp,Flanagan:2004bz,Catena:2006bd} on this issue. We favor the perspective that neither frame is more physical than the other, but that cosmological observations can be viewed more conveniently in the JF since interpretation of observations that form the key evidence for dark energy, of fluctuations in the CMB, and of redshift measurements in galaxy surveys and supernovae, are usually made with the assumption that baryons are minimally coupled. When considering the dynamical evolution of scalar-tensor theories therefore we consider properties, for example the fractional matter density, equation of state and stability of critical points (attractors/repellers), in the JF to be those that are most appropriate to compare with observational constraints. 

The Einstein frame  perspective is also useful however, in that the evolution equations and fixed point analysis are arguably easier analytically in this frame.  

The Einstein frame description can also be used to encompass theories outside of the modified gravity realm. In these theories the scalar is minimally coupled to gravity but the coupling strength of the scalar  with cold dark matter ($\beta_{c}$), baryons ($\beta_{b}$) and neutrinos ($\beta_{\nu}$) could be different, for example (\cite{Amendola:1999er, Bean:2000zm,Bean:2001ys,Fardon:2003eh,Fardon:2005wc,Afshordi:2005ym, Bjaelde:2007ki, Bean:2007nx, Bean:2007ny}). In models in which CDM alone is coupled observational constraints constrain $\beta_c < 0.1$ \cite{Amendola:1999er}.  In Chameleon Cosmology models \cite{Khoury:2003aq,Khoury:2003rn,Brax:2004qh}, the scalar fields acquire a mass whose magnitude depends on the local matter density, and all $\beta_{i}$ can be spatially varying, and of order unity. 

One of the aims of this paper is to give a coherent picture of dynamical attractor analysis in modified gravity theories in both the Einstein and Jordan frames. As will come out naturally in our analysis, the choice of frame is really a matter of convenience and the physical implications of the analysis, in terms of attractor critical points and their stability arises identically out of analyses irrespective of frame choice.
We, therefore, focus our analysis on scenarios in which the coupling strength is the same for all types of matter and in which the Jordan frame quantities are to be compared with observational predictions. The Einstein frame analysis however can be equally applied to theories in which CDM-scalar couplings are present, and the Einstein frame is the physical frame.

\subsection{The form of the scalar potential}
\label{Twod}
The action in (\ref{JFaction}) allows both a free choice of coupling to gravity $F(\Phi)$ and scalar self interaction potential $V(\Phi)$. In order to include an important group of extended gravity theories, `$f(R)$ theories', and $\Lambda$CDM  as classes of models described by our analysis, however, we place a restriction on the form of the JF potential.

$f(R)$ theories, where the Lagrangian contains an arbitrary function of the scalar curvature $R$, can be written as a scalar tensor theory of gravity \cite{Maeda:1987xf,Maeda:1988ab,Chiba:2003ir,Flanagan:2003rb},
\begin{eqnarray}
S &=& \frac{1}{2\kappa^2}\int d^4 x \sqrt{-g}f(R)+ S_{m} 
\\
&=& \frac{1}{2\kappa^2}\int d^4 x \sqrt{-g}[F(\Phi) R-(\Phi F(\Phi)-f(\Phi))]+ S_{m} 
\end{eqnarray}
where,
\begin{eqnarray}
	F(\Phi) & \equiv & \textrm{d}f(\Phi)/\textrm{d}\Phi. \label{Fdef}
\end{eqnarray}
This is equivalent to our general scalar-tensor action (\ref{JFaction}), with $\beta= 1/2$, \textit{i.e.} a non-dynamical scalar field, and scalar potential,
\begin{equation}
	V(\Phi) = \frac{1}{2\kappa^{2}} [\Phi F(\Phi) - f(\Phi)].\label{potchoice}
\end{equation} 
By considering actions with potentials of the form (\ref{potchoice}), our analysis has $f(R)$ theories as its limit when the field $\Phi$ is non-dynamical. Equally, with this form of the potential, general relativity is regained for $f(\Phi)=\Phi$ and $\beta=0$. 

In the dynamical analysis described in the following sections, with this restricted potential, we find it useful to split the potential in  (\ref{potchoice}) into two components,
\begin{eqnarray}
	V_1(\Phi)  &\equiv & - \frac{1}{2\kappa^{2}}f(\Phi), \ \ \ 
	V_2(\Phi)  \equiv\frac{1}{2\kappa^{2}} \Phi F(\Phi),
\end{eqnarray}
and consider their relative importance in the cosmic evolution. This analysis could be extended, therefore, to describe evolution of more general actions, without the restriction on the potential given in (\ref{potchoice}),  through writing
 \begin{equation}
	V(\Phi) = \frac{1}{2\kappa^{2}} [\Phi F(\Phi) - f(\Phi)] + V_{3}(\Phi)
\end{equation} 
and considering the relative importance of the extra component, $V_3(\Phi)$, to the dynamical evolution. We leave such an extension to future work.

\section{Dynamical critical points}
\label{Three}

\subsection{Jordan frame autonomous phase plane equations}
\label{Threea}

We would like to find the critical points of the dynamical background evolution. This equates to finding asymptotic power-law solutions of the form $a\propto t^{p}$, where $p$ is a constant and related to the effective equation of state of the system, $w_{eff}$, by $p=2/3(1+w_{eff})$.

Writing the scalar potential as (\ref{potchoice}), and defining the dynamical variables,
\begin{eqnarray}
	x_{1} \equiv - \frac{\dot{F}}{HF} = - \frac{d\ln F}{d\ln a}, \ \
	x_{2} \equiv - \frac{f}{6FH^{2}}, \ \
	x_{3} \equiv \frac{\Phi}{6H^{2}},
\end{eqnarray}
the fractional matter density in the JF can be expressed in terms of these dynamical parameters through the Friedmann equation (\ref{FriedmannJF}), as,
\bea
	\Omega_{m}& \equiv& \frac{\kappa^{2}\rho_{m}}{3FH^{2}} = 1 - x_{1}-K x_1^2- x_{2} -x_{3}, \label{omjf}
\eea
where $K \equiv  (1-4\beta^2)/16\beta^2$.
$w_{eff}$ can also be expressed in terms of the dynamical parameters,
\bea
	\fl w_{eff} & = & - \frac{2}{3}\frac{\dot{H}}{H^2} -1 \\ 
	\fl & = & \frac{4\beta^2}{3}+\frac{\left(1-4\beta^2\right)}{3}x_1+\left(1-\frac{4\beta^2}{3}\right)K x_1^2 - \left(1-4\beta^2\right)x_2 - \left(1 - \frac{4\beta^2}{3}\right)x_3
\eea
so that critical points, where $w_{eff}$ is constant in time, are satisfied by $x_1'=x_2'=x_3'=0$, where $x'\equiv dx/d\ln a = \dot{x}/H$.

The closed set of `autonomous phase plane' equations, $x_1'$, $x_2'$ and $x_3'$ in terms of $x_1,x_2$ and $x_3$ are obtained from the JF equations (\ref{FriedmannJF}) - (\ref{JFflu3}), and constraint equation (\ref{omjf}), 
\bea
	\fl x_1' & = & - \frac{x_{1}}{2}\left[ (3-4\beta^{2}) - (3-4\beta^{2})x_{1} - (3-4\beta^{2})Kx_{1}^{2} + (3-12\beta^{2})x_{2} + (3-4\beta^{2})x_{3} \right] \nonumber \\
	\fl	& & -4\beta^{2}(1-x_{1}-Kx_{1}^{2}+3x_{2}+x_{3}) \\
	\fl x_2' & = &  x_2\left[\left(3+4\beta^2\right)+\left(2-4\beta^2\right)x_1+\left(3-4\beta^2\right)K x_1^2-3\left(1-4\beta^2\right)x_2 - \left(3-4\beta^2\right)x_3\right] \nonumber \\
	\fl & & + \frac{x_1x_3}{m} \\
	\fl x_3' & = & x_3\left[\left(3+4\beta^2\right)+\left(1-4\beta^2\right)x_1+\left(3-4\beta^2\right)K x_1^2-3\left(1-4\beta^2\right)x_2-\left(3-4\beta^2\right)x_3\right] \nonumber \\
	\fl & & -\frac{x_1x_3}{m}
\eea
where $m$ is defined as 
\bea
	m & \equiv &\frac{d\ln F}{d\ln\Phi } = -x_1\left(\frac{d\ln x_3}{d\ln a} -3(1+w_{eff})\right)^{-1}
	\label{mdef}.
\eea
and its explicit evolution depends on the choice of $f(\Phi)$.

\subsection {Einstein frame autonomous phase plane equations}
\label{Threeb}
%
We can define analogous dimensionless dynamical variables, $\tilde{x}$, $\tilde{y}$ and $\tilde{z}$, to those defined in Sec. \ref{Threea}, which are simply related to the fractional energy densitites of the scalar kinetic and potential energies in the EF,
\begin{eqnarray}
\tilde{x} = \frac{\kappa} {\sqrt{6} \tilde{H}} \frac{d\phi}{d\tilde{t}},\ \ \; 
\tilde{y} = \frac{\kappa^{2} \tilde{V}_{1}} {3 \tilde{H}^{2}},  \ \ \; 
\tilde{z} = \frac{\kappa^{2} \tilde{V}_{2}} {3 \tilde{H}^{2}},
\end{eqnarray}
where $\tilde{V}(\phi)  =  \tilde{V}_{1}(\phi) + \tilde{V}_{2}(\phi)$,
\begin{eqnarray}
\tilde{V}_{1}(\phi)& \equiv & - \frac{f}{2\kappa^{2}F^{2}}, \ \ \ 
  \tilde{V}_{2}(\phi) \equiv  \frac{\Phi F}{2\kappa^{2}F^{2}}.
\end{eqnarray}
The specific form of the potentials $\tilde V_1$ and $\tilde V_2$  coming from the conformal transformation will depend on the form of $f(\Phi)$.

We can also define an Einstein frame fractional energy density for matter, and using (\ref{FriedmannEF}) write it in terms of the Einstein frame dynamical variables,
  \begin{equation}
	\tilde{\Omega}_{m} \equiv \frac{\kappa^{2}\tilde{\rho}_{m}}{3\tilde{H}^{2}} = 1-(\tilde{x}^{2}+\tilde{y}+\tilde{z}) \label{omef}
\end{equation}
where $0\le 	\tilde{\Omega}_{m} \le 1.$

On differentiating the potentials $\tilde{V}_{1}(\phi) $ and $\tilde{V}_{2}(\phi)$ with respect to $\phi$ and using the scalar field redefinition equation (\ref{redefinition}) we get,
\begin{eqnarray}
	\tilde{V}_{1,\phi} & = & - 2 \sqrt{\frac{2}{3}} \kappa \beta \left[ 2\tilde{V}_{1} + \frac{1}{m} \tilde{V}_{2} \right],
	 \\
	\tilde{V}_{2,\phi} & = & 2 \sqrt{\frac{2}{3}} \kappa \beta \tilde{V}_{2} \left[ \frac{1}{m} - 1 \right].
\end{eqnarray}
 
The EF equations (\ref{FriedmannEF}) - (\ref{bkg}), along with the constraint equation (\ref{omef}) can be used to write down the system of plane-autonomous equations,
\begin{eqnarray}
	\frac{d\tilde{x}}{d\ln \tilde{a}} & = & \beta(1-\tilde{x}^{2}-\tilde{y}-\tilde{z}) + 2\beta(2\tilde{y}+\tilde{z}) - \frac{3}{2}\tilde{x}(1-\tilde{x}^{2}+\tilde{y}+\tilde{z}), \\
	\frac{d\tilde{y}}{d\ln \tilde{a}}  & = & - 4\beta \tilde{x}\tilde{z} \left( \frac{1}{m} \right) - 8\beta \tilde{x}\tilde{y} + 3\tilde{y}(1+\tilde{x}^{2}-\tilde{y}-\tilde{z}), \\
	\frac{d\tilde{z}}{d\ln \tilde{a}}  & = & 4\beta \tilde{x}\tilde{z} \left( \frac{1}{m}-1 \right) + 3\tilde{z}(1+\tilde{x}^{2}-\tilde{y}-\tilde{z}).
\end{eqnarray}
where all information about the scalar field redefinition $\phi(\Phi)$ is encoded in the parameter $m$.

We can find EF evolution where there is power law expansion of the form $\tilde{a}\propto \tilde{t}^{\tilde{p}}$ where $\tilde{p}$ is constant by solving $d\tilde{x}/d\ln \tilde{a}=d\tilde{y}/d\ln \tilde{a}=d\tilde{z}/d\ln \tilde{a}=0$. As discussed previously, if visible matter, and not only CDM,  is non-minimally coupled in the Einstein frame then Einstein frame is not the one in which cosmological observations are usually expressed. However as we discuss in the following section, a mapping exists between the autonomous phase planes in each frame that implies that solving one set of equations translates into the solutions of the other.

\subsection {Mapping between the Einstein and Jordan frame phase planes}
\label{Threec}
There is a direct mapping between the JF and EF variables,
\bea
x_1 &=&-\frac{4\beta \tilde{x}}{1-2\beta\tilde{x}},  \ \ \ \  \tilde{x} =\frac{1}{2\beta}\frac{x_1}{x_1-2}, \label{xdef}
\\
x_{2} & = & \frac{\tilde{y}}{(1 - 2\beta \tilde{x})^{2}}, \ \ \ \  \tilde{y} = \frac{4x_2}{(x_1-2)^2}, \label{ydef}
\\
x_{3} & = & \frac{\tilde{z}}{(1 - 2\beta \tilde{x})^{2}}, \ \ \ \ \tilde{z}= \frac{4x_3}{(x_1-2)^2},\label{zdef}
\eea
where 
\begin{equation}
	\frac{d \ln \tilde{a}}{d \ln a} = \frac{1}{1 - 2\beta \tilde{x}}=-\frac{x_1-2}{2}.
	\label{transfactor}
\end{equation} 

 Writing a point in the JF as $\mathbf{x} \equiv (x_{1}, x_{2}, x_{3})$  and a point in the EF as $\mathbf{\tilde{x}} \equiv (\tilde{x}, \tilde{y}, \tilde{z})$, the transformation matrix $\mathcal{T}$ defined such that $\mathbf{\Delta x}= \mathcal{T} \mathbf{\Delta\tilde{x}}$,  where $\mathbf{\Delta{x}} \equiv( \Delta x_1 ,  \Delta x_2, \Delta x_3)$ and $\mathbf{\Delta \tilde{x}} \equiv( \Delta \tilde{x} ,  \Delta \tilde{y} , \Delta \tilde{z})$, is given by, 
\begin{eqnarray}
	\mathcal{T} = 
	\frac{1}{(1 - 2\beta \tilde{x})^{2}} 
	\left(
	\begin{array}{c c c}
		-4\beta & 0 & 0 \\
		\frac{4\beta \tilde{y}}{1 - 2\beta \tilde{x}} & 1 & 0 \\
		\frac{4\beta \tilde{z}}{1 - 2\beta \tilde{x}} & 0 & 1
	\end{array}  
	\right),
	\label{Tdef}
\end{eqnarray}

\begin{eqnarray}
	\mathcal{T}^{-1} = 
	\frac{1}{(x_1-2)^{2}} 
	\left(
	\begin{array}{c c c}
		-\frac{1}{\beta}& 0 & 0 \\
		-\frac{8x_2}{(x_1-2)} & 4 & 0 \\
		-\frac{8x_3}{(x_1-2)} & 0 & 4
	\end{array} 
	\right). \label{Tinvdef}
\end{eqnarray}

The transformation from the Jordan to Einstein frame is well defined as long as $x_1\neq 2$. Equation (\ref{xdef}) shows that for $x_1\neq 2$, if follows that $(1-2\beta\tilde{x})\neq 0 $ and the transformation from the Einstein to Jordan frame is non-singular. For theories in which $x_1=2$ occurs, this simply represents that the Einstein frame mapping is ill-defined and the dynamical analysis should be undertaken in the Jordan frame. 

\subsection {Critical points and calculating their stability}
\label{Threed}
The critical points (also called fixed points) in the Jordan and Einstein frames, $\mathbf{x_c} \equiv (x_{1c}, x_{2c}, x_{3c})$  and $\mathbf{\tilde{x}_{c}}\equiv(\tilde{x}_{c}, \tilde{y}_{c}, \tilde{z}_{c})$ respectively, are the solutions of $x_1'=x_2'=x_3'=0$  and $d\tilde{x}/d\ln\tilde{a}=d\tilde{y}/d\ln\tilde{a}=d\tilde{z}/d\ln\tilde{a}=0$. These points may be stable, saddle, or unstable solutions. In order to study the stability of the critical points  we expand about these points, $\mathbf{x}\equiv \mathbf{x_{c}}+\mathbf{\Delta x}$ and  $\mathbf{\tilde{x}}\equiv \mathbf{\tilde{x}_{c}}+\mathbf{\Delta\tilde{x}}$ 
and consider the eigenvalues of the stability matrices ${\mathcal M}$ and $\tilde{\mathcal M}$ defined by $\mathbf{\Delta x}'\equiv {\mathcal{M}}\mathbf{\Delta x}$,
and  $d \mathbf{\Delta \tilde{x}}/d \ln  \tilde{a}\equiv  \tilde{\mathcal{M}}\mathbf{\Delta\tilde{x}}$. 

For any point the matrices  ${\mathcal{M}}$ and $ \tilde{\mathcal{M}}$ are related by
\bea
 {\mathcal{M}} =  \frac{d\ln \tilde a}{d\ln a}\left[{\cal T}\tilde{\cal M}{\cal T}^{-1}+\frac{d ({\cal T})}{d\ln \tilde a}{\cal T}^{-1}\right]
\eea
From (\ref{transfactor}), for well-defined transformations ($x_1\ne 2$), at the critical points $d ({\cal T})/d\ln \tilde a =0$ so that
we can relate $\mathcal{M}$, $\tilde{\mathcal{M}}$ and $\mathcal{T}$,
\begin{equation}
	\mathcal{M} = \frac{1}{1 - 2\beta \tilde{x}_{c}} \mathcal{T}_{c} \tilde{\mathcal{M}} \mathcal{T}_{c}^{-1}. 
	\label{simtrans}
\end{equation}
where $\mathcal{T}_{c}$ is the transformation matrix at the critical point.

Let the eigenvalues of $\mathcal{M}$ at some critical point be $\lambda_{1}$, $\lambda_{2}$ and $\lambda_{3}$. Then the general solution for the evolution of linear perturbations in the JF will be of the form,
\begin{eqnarray}
	\mathbf{\Delta x} = \mathbf{c_{1}}e^{\lambda_{1} \Delta N} + \mathbf{c_{2}}e^{\lambda_{2} \Delta N} + \mathbf{c_{3}}e^{\lambda_{3} \Delta N}
\end{eqnarray}
where $\mathbf{c_{1}}$, $\mathbf{c_{2}}$, and $\mathbf{c_{3}}$ are vectors of constants, and $N=\ln a$.  We require the linear perturbations to decay with time so that the critical point is stable. Cosmological observations of the cosmic expansion, as measured by the CMB, supernovae and galaxy surveys are consistent with a monotonically increasing expansion history such as in $\Lambda$CDM \cite{Perlmutter:1998np}-\cite{Kitching:2006mq}. We therefore consider stability criteria as the universe expands, taking $\Delta N$ positive, in which case stable fixed points are characterized by negative real parts of eigenvalues in the Jordan frame. If the universe had been contracting instead of expanding at any point of time, then corresponding fixed points would map to repellers instead of attractors.

If the eigenvalues of $\tilde{\mathcal{M}}$ at some critical point $(\tilde{x}_{c}, \tilde{y}_{c}, \tilde{z}_{c})$ are $\tilde{\lambda}_{1}$, $\tilde{\lambda}_{2}$ and $\tilde{\lambda}_{3}$, then the general solution for the evolution of linear perturbations in the EF will be of the form,
\begin{eqnarray}
	\mathbf{\Delta \tilde{x}} = \mathbf{\tilde{c}_{1}}e^{\tilde{\lambda}_{1} \Delta \tilde{N}} + \mathbf{\tilde{c}_{2}}e^{\tilde{\lambda}_{2} \Delta \tilde{N}} + \mathbf{\tilde{c}_{3}}e^{\tilde{\lambda}_{3} \Delta \tilde{N}}
\end{eqnarray}
where $\mathbf{\tilde{c}_{1}}$, $\mathbf{\tilde{c}_{2}}$, and $\mathbf{\tilde{c}_{3}}$ are vectors of constants and $\tilde{N} = \ln \tilde{a}$.  We require the linear perturbations to decay with time so that the critical point is stable.  Our time variable in the EF, $\tilde{N}$, may increase or decrease with the evolution of the universe since at a critical point, $\Delta N = (1-2\beta \tilde{x}_{c}) \Delta \tilde{N}$ and $(1-2\beta\tilde{x}_{c})$ could be positive or negative. However the eigenvalues in the two frames at a critical point are related by $\mathbf{\lambda} = \mathbf{\tilde{\lambda}}/(1-2\beta \tilde{x}_{c})$. The product of these two together hence guarantees that stability in one frame implies stability in the other frame irrespective of the sign of $(1-2\beta \tilde{x}_{c})$. The sign of the stable eigenvalue in the Einstein frame may therefore be opposite to that in the Jordan frame.

The stability matrix in the Jordan frame, ${\mathcal{M}}$, is given by:
\begin{eqnarray}
	{\mathcal{M}}_{11} & = & 2\beta^{2}(3-2x_{1c}+4Kx_{1c}-3Kx_{1c}^{2}+3x_{2c}+x_{3c}) \nonumber 
	\\ && - \frac{3}{2}(1-2x_{1c}-3Kx_{1c}^{2}+x_{2c}+x_{3c}), \ \ \ \ \ \\
	{\mathcal{M}}_{12} & = & -12\beta^{2} - \frac{3}{2}x_{1c} + 6\beta^{2}x_{1c}, \\
	{\mathcal{M}}_{13} & = & -4\beta^{2} - \frac{3}{2}x_{1c} + 2\beta^{2}x_{1c}, \\
	{\mathcal{M}}_{21} & = & \frac{x_{3c}}{m} + (2-4\beta^{2})x_{2c} + 2K(3-4\beta^{2})x_{1c}x_{2c}, \\
	{\mathcal{M}}_{22} & = & 3+4\beta^{2} + (2-4\beta^{2})x_{1c} + \frac{m_{r}r^{2}x_{1c}}{m^{2}} + (3-4\beta^{2})Kx_{1c}^{2} \nonumber 
	\\ &&- 6(1-4\beta^{2})x_{2c}  - (3-4\beta^{2})x_{3c}, \ \ \ \ \\
	{\mathcal{M}}_{23} & = & \frac{x_{1c}}{m} - \frac{m_{r}rx_{1c}}{m^{2}} - (3-4\beta^{2})x_{2c}, \\
	{\mathcal{M}}_{31} & = & \frac{-x_{3c}}{m} + (1-4\beta^{2})x_{3c} + 2K(3-4\beta^{2})x_{1c}x_{3c}, \\
	{\mathcal{M}}_{32} & = & \frac{-m_{r}r^{2}x_{1c}}{m^{2}} - 3(1-4\beta^{2})x_{3c}, \\
	{\mathcal{M}}_{33} & = & 3+4\beta^{2} + (1-4\beta^{2})x_{1c} - \frac{x_{1c}}{m} + \frac{m_{r}rx_{1c}}{m^{2}} + (3-4\beta^{2})Kx_{1c}^{2} \nonumber 
	\\ && - 3(1-4\beta^{2})x_{2c} - 2(3-4\beta^{2})x_{3c}. \ \ \ \ \ 
\end{eqnarray} 
where $m_{r} \equiv dm/dr$, and, 
\begin{eqnarray}
	r & \equiv & - \frac{\Phi F}{f} = - \frac{d\ln f}{d\ln \Phi}\; = \; \frac{x_{3}}{x_{2}}\; = \; \frac{\tilde{z}}{\tilde{y}} \label{rdef}.
\end{eqnarray}

The stability matrix in the Einstein frame, $\tilde{\mathcal{M}}$, is given by:
\begin{eqnarray}
	\tilde{\mathcal{M}}_{11} & = & - 2\beta \tilde{x}_{c} + 3\tilde{x}_{c}^{2} - \frac{3}{2}(1-\tilde{x}_{c}^{2}+\tilde{y}_{c}+\tilde{z}_{c}), \\
	\tilde{\mathcal{M}}_{12} & = & 3\beta - \frac{3}{2}\tilde{x}_{c}, \\
	\tilde{\mathcal{M}}_{13} & = & \beta - \frac{3}{2}\tilde{x}_{c}, \\
	\tilde{\mathcal{M}}_{21} & = & \frac{-4\beta \tilde{z}_{c}}{m} - 8\beta \tilde{y}_{c} + 6\tilde{x}_{c}\tilde{y}_{c}, \\
	\tilde{\mathcal{M}}_{22} & = & -8\beta \tilde{x}_{c} - \frac{4\beta m_{r} r^{2} \tilde{x}_{c}}{m^{2}} + 3(1+\tilde{x}_{c}^{2}-\tilde{y}_{c}-\tilde{z}_{c}) - 3\tilde{y}_{c}, \\
	\tilde{\mathcal{M}}_{23} & = & \frac{-4\beta \tilde{x}_{c}}{m} + \frac{4\beta m_{r} r \tilde{x}_{c}}{m^{2}} - 3\tilde{y}_{c}, \\
	\tilde{\mathcal{M}}_{31} & = & 6\tilde{x}_{c}\tilde{z}_{c} + 4\beta \tilde{z}_{c}\left( -1+\frac{1}{m} \right), \\
	\tilde{\mathcal{M}}_{32} & = & \frac{4\beta m_{r} r^{2} \tilde{x}_{c}}{m^{2}} - 3\tilde{z}_{c}, \\
	\tilde{\mathcal{M}}_{33} & = & 4\beta \tilde{x}_{c}\left( -1+\frac{1}{m} \right) - \frac{4\beta m_{r} r \tilde{x}_{c}}{m^{2}} - 3\tilde{z}_{c} +  3(1+\tilde{x}_{c}^{2}-\tilde{y}_{c}-\tilde{z}_{c}) 
\end{eqnarray} 

\section{Solving the autonomous phase plane equations}
\label{Four}
\subsection{The critical points}
\label{Foura}

The seven critical points, $\mathbbm{P}1-\mathbbm{P}7$,  are given in Table \ref{table:critpointsEF} and Table \ref{table:critpointsJF}  in the Einstein and Jordan frames respectively. Table \ref{table:observables} gives the Jordan frame fractional matter density $\Omega_{m}$ and effective equation of state parameter $w_{eff}$ for each of the points, which can be compared to observational constraints.

\begin{table}[t]
\centering
\caption{Critical points ($\mathbbm{P}1-\mathbbm{P}7$) of the dynamical attractors for a universal, general scalar coupling, $\beta$ in the Einstein frame in terms of dynamical variables $(\tilde{x},\tilde{y},\tilde{z})$} 
{\small
\begin{tabular}{c c c c} \\
\hline\hline \\
Point & $\tilde{x}$ & $\tilde{y}$ & $\tilde{z}$ 
\\ \\
\hline \\
$\mathbbm{P}1$  & 0 & -1 & 2
\\ \\
$\mathbbm{P}2$  & $\frac{2\beta}{3}$ & 0 & 0 
\\ \\
$\mathbbm{P}3$  & -1 & 0 & 0 
\\ \\
$\mathbbm{P}4$  & $\frac{4\beta}{3}$ & $\frac{9-16\beta^{2}}{9}$ & 0 
\\ \\
$\mathbbm{P}5$  & $\frac{3m}{-4\beta + 2\beta m}$ & $\frac{-8\beta^{2} + m(-9+4\beta^{2})} {4\beta^{2}(-2+m)^{2}}$ & $\frac{-(1+m)[-8\beta^{2} + m(-9+4\beta^{2})]} {4\beta^{2}(-2+m)^{2}}$ 
\\ \\
$\mathbbm{P}6$  & $\frac{2\beta(-1+m)}{3m}$ & $\frac{4\beta^{2} - 8\beta^{2}m + m^{2}(-9+4\beta^{2})}{9m^{3}}$ & $\frac{-(1+m)[4\beta^{2} - 8\beta^{2}m + m^{2}(-9+4\beta^{2})]}{9m^{3}}$ 
\\ \\
$\mathbbm{P}7$  & 1 & 0 & 0
\\ \\
\hline\hline\\
\end{tabular}}
\label{table:critpointsEF}
\end{table}
\begin{table}[t]
\centering
\caption{Critical points ($\mathbbm{P}1-\mathbbm{P}7$) of the dynamical attractors for a universal, general scalar coupling, $\beta$ in the Jordan frame in terms of dynamical variables $(x_{1},x_{2},x_{3})$} 
{\small
\begin{tabular}{c c c c} \\
\hline\hline \\
Point & $x_{1}$ & $x_{2}$ & $x_{3}$ 
\\ \\
\hline \\
$\mathbbm{P}1$  & 0 & -1 & 2
\\ \\
$\mathbbm{P}2$  & $\frac{8\beta^2}{4\beta^2-3}$& 0 & 0
\\ \\
$\mathbbm{P}3$  &$\frac{4\beta}{2\beta+1}$ & 0 & 0 
\\ \\
$\mathbbm{P}4$  &$\frac{16\beta^2}{8\beta^2-3}$& $\frac{9-16\beta^{2}}{(3-8\beta^{2})^{2}}$ & 0
\\ \\
$\mathbbm{P}5$  & $\frac{3m}{1+m}$ & $\frac{-9m+4\beta^{2}(-2+m)}{16\beta^{2}(1+m)^{2}}$ & $\frac{9m-4\beta^{2}(-2+m)}{16\beta^{2}(1+m)}$ 
\\ \\
$\mathbbm{P}6$  & $\frac{8\beta^{2}(1-m)}{3m+4\beta^{2}(1-m)}$ & $\frac{-9m^{2}+4\beta^{2}(-1+m)^{2}}{m[3m-4\beta^{2}(-1+m)]^{2}}$ & $\frac{(1+m)[9m^{2}-4\beta^{2}(-1+m)^{2}]}{m[3m-4\beta^{2}(-1+m)]^{2}}$
\\ \\
$\mathbbm{P}7$  &$\frac{4\beta}{2\beta-1}$& 0 & 0
\\ \\
\hline\hline\\
\end{tabular}}
\label{table:critpointsJF}
\end{table}

\begin{table}[t]
\centering
\caption{The fractional matter density $\Omega_m$ and effective equation of state $w_{eff}$ in the Jordan frame (JF). For modified gravity theories, observational constraints derived from redshift measurements are more directly applicable in the Jordan rather than Einstein frame, as discussed in section \ref{Twoc}.}  
{\small
\begin{tabular}{c c c} \\
\hline\hline \\
Point & $\Omega_{m} (JF)$ & $w_{eff} (JF)$ 
\\ \\
\hline \\
$\mathbbm{P}1$  & 0 & -1 
\\ \\
$\mathbbm{P}2$  &$\frac{9-4\beta^2}{(3-4\beta^2)^2}$  & $\frac{8\beta^{2}}{9-12\beta^{2}}$ 
\\ \\
$\mathbbm{P}3$  & 0 & $\frac{3-2\beta}{3+6\beta}$ 
\\ \\
$\mathbbm{P}4$  & 0 & $\frac{9-40\beta^{2}}{-9+24\beta^{2}}$ 
\\ \\
$\mathbbm{P}5$  & $\frac{-9m^{2}+4\beta^{2}(2-3m+m^2)}{8\beta^{2}(1+m)^{2}}$ & $\frac{-m}{1+m}$ 
\\ \\
$\mathbbm{P}6$  & 0 & $\frac{-8\beta^{2}+20\beta^{2}m+m^{2}(9-12\beta^{2})}{3m[-4\beta^{2}+m(-3+4\beta^{2})]}$ 
\\ \\
$\mathbbm{P}7$  & 0 & $\frac{3+2\beta}{3-6\beta}$
\\ \\
\hline\hline\\
\end{tabular}}
\label{table:observables}
\end{table}

We will find that $r$ and $m$, which we can treat as  dynamical variables determined by the specific theory (the specific form of $f(\Phi)$), are useful parameters to characterize the properties and stability of the critical points.

Note that at the critical points, given that $x_{1}'=x_{2}'=x_3'=0$ and (\ref{rdef}),
\bea
r'= -rx_1\left (\frac{r+1}{m}+1\right) =0.
\eea  
\textit{i.e.} $r$ is constant at the fixed points.

At the points $\mathbbm{P}5$ and $\mathbbm{P}6$ when $x_{1c}\neq 0$ and $r\neq 0$, $m$ and $r$ are related by
\bea
m = -(r+1) \label{mrrelation}.
\eea
From (\ref{mdef}) the value of $m$ at these critical points is related to the equation of state by
\bea
m &=&  \frac{x_{1c}}{3(1+w_{eff})}\label{mwrelation}.
\eea

At $\mathbbm{P}1$, $r=-2$ irrespective of the value of $\beta$ or the form of $f(\Phi)$, however since $w_{eff}=-1$, from (\ref{mdef}), the coordinates are independent of the value of $m$. Similarly at $\mathbbm{P}4$, $r=0$ and since $x_{3c}=0$, from (\ref{mdef}), again the coordinates are independent of the value of $m$.

Finally, at points $\mathbbm{P}2$, $\mathbbm{P}3$ and $\mathbbm{P}$7, where $x_{2c}=x_{3c}=0$, the coordinates $\{x_{1},x_{2},x_{3}\}$ are not explicitly dependent on the specific values of $r$, $m$. However, as discussed in Sec. \ref{Fourc} the stability criteria / eigenvalues at these points are. Note that $r$ remains well-defined as given in (\ref{rdef}) despite $x_3$ and $x_2$, or $\tilde{z}$ and $\tilde{y}$, asympotically tending to zero at a fixed point. 

It should be noted that a specific choice of $f(\Phi)$ may only be able to reach a subset of fixed points given in this general analysis. For example, consider a scenario in which $f(\Phi)\propto \Phi^{n}$ ($n \neq 0$), where $n$ is constant, then $r=-n$ at all times, even as $\Phi\rightarrow 0$ as at $\mathbbm{P}2$, $\mathbbm{P}3$ and $\mathbbm{P}7$. This theory could reach $\mathbbm{P}1$ only if $n=-2$, and $\mathbbm{P}4$ only if $n=0$ (for which $F(\Phi)=0$ and would not give a well defined action). Moreover, the consistency relation for $\mathbbm{P}5$ and $\mathbbm{P}6$ in (\ref{mrrelation}) is always satisfied since $m=n-1$ for this particular choice of $f(\Phi)$.

We briefly summarize the properties of each fixed point below. In each case $\Omega_m$ and $w_{eff}$ are the Jordan frame quantities most easily compared with cosmological observations, as discussed in section \ref{Twoc}.

\vspace{0.25cm}
\noindent\underline {Point $\mathbbm{P}1$:}
Since $\Omega_{m}=0$, and $w_{eff}=-1$ the point $\mathbbm{P}1$ can only represent a de-Sitter accelerated universe.

 \vspace{0.25cm}
\noindent\underline {Point $\mathbbm{P}2$}: $\mathbbm{P}2$ is a generalization of the ``$\phi$ matter-dominated era'' ($\phi$MDE) attractor in $f(R)$ theories in which the scale factor in the JF has wrong time evolution, $a \propto t^{1/2}$ \cite{Amendola:2006kh}. However for general values of $\beta$, $w_{eff}$ can be arbitrarily close to zero and a good approximation to the standard MDE in which $a \propto t^{2/3}$. $\mathbbm{P}2$ could therefore act as an approximate CDM era which, for small $\beta$, could be consistent with cosmological observations. For $|\beta|>\sqrt{3}/2$, $\mathbbm{P}2$ gives accelerated expansion..
 
\vspace{0.25cm}
\noindent\underline {Point $\mathbbm{P}3$:} Since $\Omega_{m}=0$ at $\mathbbm{P}3$, it cannot represent a matter-dominated era (MDE). $\mathbbm{P}3$ describes an accelerated universe for $\beta < -1/2$. 

\begin{figure}[t]
\centering{
\includegraphics[width=9cm]{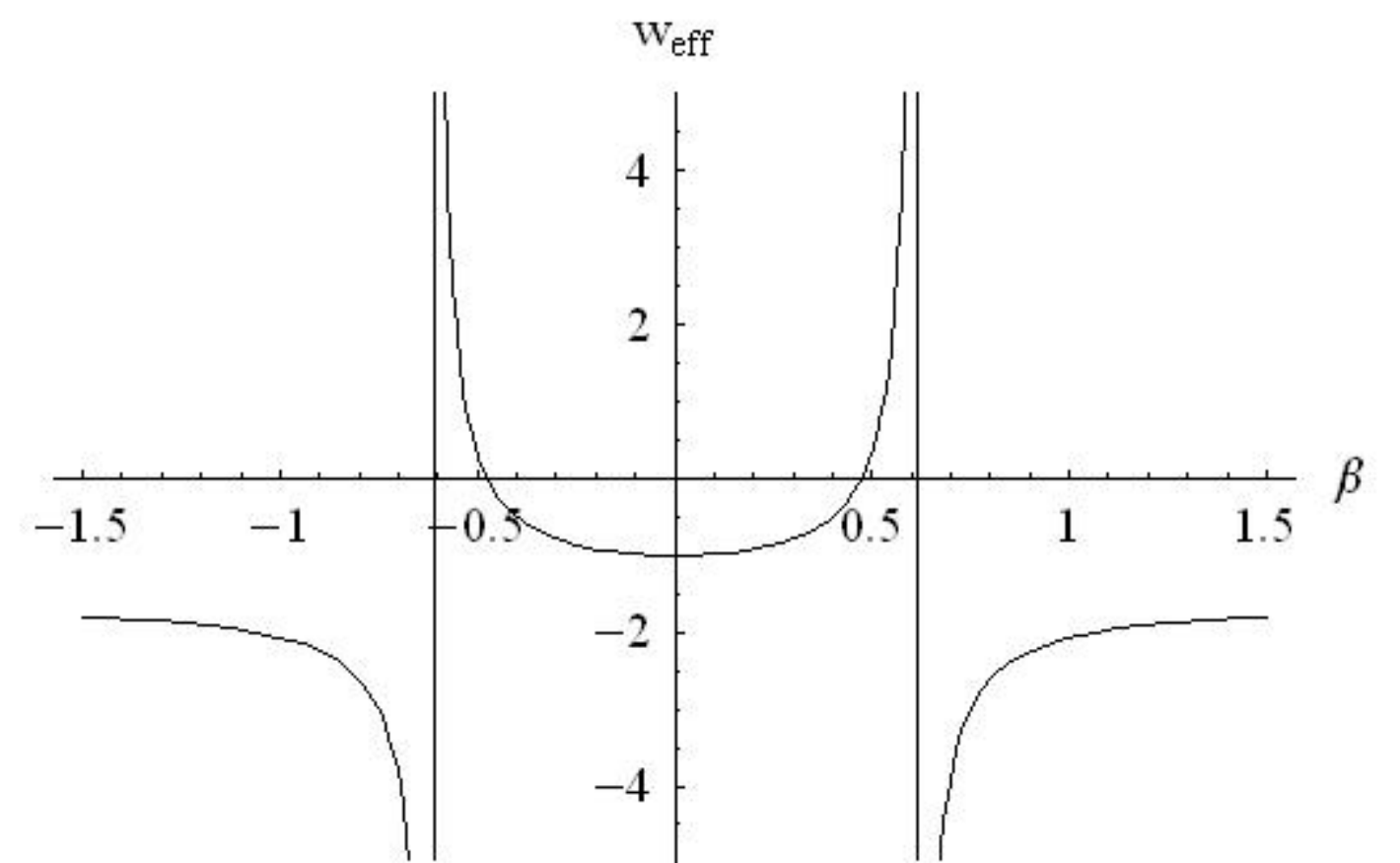}}
\caption{The effective equation of state $w_{eff}$ as a function of $\beta$ for the point $\mathbbm{P}4$.}
\label{graph1}
\end{figure}

\vspace{0.25cm}
\noindent\underline {Point $\mathbbm{P}4$:}  For $\mathbbm{P}4$ we again have $\Omega_{m}=0$, so it cannot represent a valid MDE. In Figure \ref{graph1} we show the variation of $w_{eff}$ with $\beta$ for this point. $\mathbbm{P}4$  gives rise to acceleration for $\beta$ in the range $\beta < -\sqrt{3/8}, \; \beta > \sqrt{3/8}$, or $|\beta| < \sqrt{3}/4$.   Acceleration has $w_{eff}<-1$ for $\beta \leq -3/4$ and $\beta \geq 3/4$, and $w_{eff}>-1$ for $|\beta| < \sqrt{3}/4$.   For $\beta=1/2$, $w_{eff}=1/3$, therefore $\mathbbm{P}4$ is not accelerated in $f(R)$ theories and general values of $\beta$ thus open up avenues for new accelerated critical points.  

\vspace{0.25cm}
\noindent\underline {Point $\mathbbm{P}5$:} As  $m(\mathbbm{P}5) \rightarrow 0$, $\Omega_{m}\rightarrow1$ and $w_{eff}\rightarrow 0$ for all values of $\beta$. Thus $m(r=-1)=0$ represents a standard MDE. In order for $\mathbbm{P}5$ to represent accelerated expansion, $m(\mathbbm{P}5)<-1$ or $m(\mathbbm{P}5)>1/2$. 

\vspace{0.25cm}
\noindent\underline {Point $\mathbbm{P}6$:} Since $\Omega_{m} = 0$ for $\mathbbm{P}6$, it can not be used to represent a MDE. $\mathbbm{P}6$ gives accelerated expansion for a wide range of \{$m,\beta$\}. In the limit $m(\mathbbm{P}6) \rightarrow \pm \infty$ the point $\mathbbm{P}6$ tends to a de-Sitter point with $w_{eff} \rightarrow -1$. It is also de-Sitter for $m(\mathbbm{P}6)=1$, which is the same as the point $\mathbbm{P}1$.  As $m\rightarrow 1/(1-3/4\beta^2)$ or $m\rightarrow0^{-}$, one would get  acceleration with $w_{eff}\rightarrow -\infty$. 

\vspace{0.25cm}
\noindent\underline {Point $\mathbbm{P}7$:} Since $\Omega_{m}=0$, $\mathbbm{P}7$ cannot represent a MDE. For it to represent an accelerated universe one requires $\beta > 1/2$.

\subsection{Comparison with other dynamical analysis of $f(R)$ and coupled CDM theories}
\label{Fourb}
Before discussing the nature of the critical points in detail (which we do in the next section) it is useful to make a connection between the solutions that we have found and those that have been obtained in the literature, since our general analysis maps to specific models that have been considered previously. 

Solutions for $f(R)$ models with $\beta=1/2$ have been obtained in \cite{Amendola:2006we} in the JF.  We can map solutions obtained in \cite{Amendola:2006we} to subsets of solutions found in this paper. A summary of the relationship between solutions found here and those in non-minimally coupled dark matter and $f(R)$ models is given in Table \ref{tab2}. 

Our Einstein frame analysis can be applied to CDM coupled dark energy models, with $\beta_{b}=0$, $\beta_{CDM}\ne 0$, \textit{e.g.} \cite{Copeland:1997et,Amendola:1999er, Bean:2000zm,Bean:2001ys}. Attractors in these theories have been considered with an  exponential potential in the EF,  $\tilde{V}(\phi) = A e^{\sqrt{2/3} \kappa \mu \phi}$ \cite{Copeland:1997et, Amendola:1999er}.

Only two critical points, which represent a subset of solutions of $\mathbbm{P}5$ and $\mathbbm{P}6$ with $\tilde{V}_1\propto \tilde{V}_2$, depend on $\mu$. 
For these points
\begin{equation}
	m = -(1+r) = \frac{2\beta}{\mu+2\beta}. 
\end{equation}

We consider in this work a purely scalar-tensor theory of gravity, with no higher order metric derivatives, such as the Ricci tensor, $R_{\mu\nu}$ or Riemann tensor, $R_{\alpha\beta\mu\nu}$. Modified gravity theories including such terms have been shown to be able to generate cosmic acceleration, for example Gauss-Bonnet gravity $f(G)$ where $G \equiv R^{2} - 4R^{\mu\nu}R_{\mu\nu} + R^{\alpha\beta\mu\nu}R_{\alpha\beta\mu\nu}$ (which is motivated by effective low-energy actions in string theory) \cite{Carroll:2004de,Easson:2004fq, Amendola:2005cr,Nojiri:2005jg, Nojiri:2005vv, Nojiri:2005am, Carter:2005fu,Cognola:2006eg,  Cognola:2006sp, Cognola:2007vq, Li:2007jm}. Our analysis is not applicable to higher derivative gravity theories.  Dynamical analyses  \cite{Amendola:2005cr,Li:2007jm} have been used to show, however,  that $f(G)$ models are highly constrained by cosmological data. 
   
\begin{table}[t]
\centering
\caption{Table showing how theories with purely CDM couplings in which the analysis is wholly in the Einstein frame, and $f(R)$ theories with $\beta=1/2$, can be mapped to subsets of the attractor behavior with a general coupling, $\beta$.} 
\vspace{0.25cm}\label{tab2}

\begin{tabular}{|c| c| c| c|} 
\hline\hline &&
\\
Point & CDM coupling, $\beta_b=0$, $\beta_c=\beta$, \cite{Amendola:1999er}  &$f(R)$ theories, $\beta=1/2$, \cite{Amendola:2006we}
\\  &&
\\ \hline 
$\mathbbm{P}1$  & - & Point $P_{1}$ \\ 
$\mathbbm{P}2$  & Point $c_{M}$ & Point $P_{2}$ \\ 
$\mathbbm{P}3$  & Point $d$ & Point $P_{3}$ \\ 
$\mathbbm{P}4$  & - & Point $P_{4}$ \\ 
$\mathbbm{P}5$ & Point $b_{M}$ (for constant $r$ and $m$) & Point $P_{5}$ \\
$\mathbbm{P}6$ & Point $a$ (for constant $r$ and $m$) & Point $P_{6}$ \\
$\mathbbm{P}7$  & Point $e$ & - \\ 
\hline\hline
\end{tabular}
\label{table:mapping}
\end{table}

\subsection{Stability criteria for the dynamical attractors}
\label{Fourc}

We present here the criteria for stability in each of the critical points discussed in Sec. \ref{Foura} and shown in Tables \ref{table:critpointsEF} and \ref{table:critpointsJF}. A viable cosmology requires an extended matter-dominated era (MDE) followed by late time acceleration in the Jordan frame. Therfore the MDE needs to be a saddle point and the acceleration can be stable or saddle. The stability criteria presented here is for an expanding universe. In the case of contraction, stable attractors map to unstable repellers and vice versa.

In Sec. \ref{Foura} we discussed the dependency of the observables $\Omega_m$ and $w_{eff}$ on $m$ and $\beta$. At certain critical points,  $\mathbbm{P}1$, $\mathbbm{P}4-\mathbbm{P}6$, $r$ is a constant or equal to $-(m+1)$, while for the points, $\mathbbm{P}2$, $\mathbbm{P}3$ and $\mathbbm{P}7$,  $r$  is unconstrained.  Here we find that the conditions for saddle matter domination and stable acceleration can be expressed in terms of conditions on $m$, $r$ and the coupling, $\beta$ for all points $\mathbbm{P}1-\mathbbm{P}7$.  

$m$ and $r$ are considered variables determined by the specific theory, in terms of which the stability of the critical points is described. We denote the value of $m$ at the critical point $\mathbbm{P}i$ by $m_{i}\equiv m(\mathbbm{P}i)$ and the derivative of $m_{i}$ with respect to $r$ at that point by $m_{ir}\equiv dm/dr(\mathbbm{P}i)$. In this section we consider conditions to give saddle MDE and stable acceleration in the Jordan frame. We then give conditions on the specific evolutionary paths for $m$ and $r$ that yield viable cosmologies in Sec. \ref{Fourd}, and discuss conditions for saddle acceleration  in Sec. \ref{Foure}. 


\vspace{0.25cm}
\noindent\underline {Point $\mathbbm{P}1$:} 

\noindent The eigenvalues for $\mathbbm{P}1$ are,
\begin{equation*}
-3, \; -\frac{3}{2} \pm \frac{1}{2} \sqrt{9 + 64 \beta^{2} \left( 1-\frac{1}{m_{1}} \right)}
\end{equation*}
where $m_{1} = m(r=-2)$.  For the point to be stable we need the real parts of all 3 eigenvalues to be 	negative (or zero). This amounts to $0 < m(r=-2) \leq 1$ irrespective of the value of $\beta$. $\mathbbm{P}1$ is a saddle point otherwise. 

			
\vspace{0.25cm}
\noindent\underline {Point $\mathbbm{P}2$:}
 
\noindent The eigenvalues for $\mathbbm{P}2$ are,
\begin{eqnarray}
	\fl -\frac{1}{2} - \frac{3}{3-4\beta^{2}}, \ \ \frac{1}{m_{2}^{2}(3-4\beta^{2})} \left.\Bigg[ 9m_{2}^{2} - 4\beta^{2} \left\{m_{2}(-1+2m_{2})+m_{2r}r(1+r) \right\} \right. \nonumber \\
	\fl \ \ \left.\pm 4\beta^{2}	\sqrt{m_{2}^{4}+2m_{2}^{3}+m_{2}^{2}\left\{1+2m_{2r}r(-1+r)\right\}-2m_{2} m_{2r}r(1+r)+m_{2r}^{2}r^{2}(1+r)^{2}} \right.\Bigg]. \nonumber
\end{eqnarray}
As an illustrative example, if $m_{2}$ is constant so that $m_{2r}=0$, the eigenvalues reduce to,
\begin{equation*}
-\frac{1}{2} - \frac{3}{3-4\beta^{2}}, \; 3, \; \frac{9m_{2}+4\beta^{2}(2-m_{2})}{m_{2}(3-4\beta^{2})}
\end{equation*}
Since at least one eigenvalue is positive, $\mathbbm{P}2$ can either be a saddle point or unstable and does not represent a stable acceleration point even though $w_{eff}<-1/3$ for $|\beta|>\sqrt{3}/2$.

			 
\vspace{0.25cm}
\noindent\underline {Point $\mathbbm{P}3$:}

\noindent The eigenvalues for $\mathbbm{P}3$ are,
\begin{eqnarray}
	\fl \frac{3+2\beta}{1+2\beta}, \ \ \frac{2}{m_{3}^{2}(1+2\beta)} \left.\Bigg[ -\beta m_{3} + \beta m_{3r}r(1+r) + 3m_{3}^{2}(1+\beta) \right. \nonumber \\
  \fl \ \ \left.\pm \beta \sqrt{m_{3}^{4}+2m_{3}^{3}+m_{3}^{2}\left\{1+2m_{3r}r(-1+r)\right\}-2m_{3} m_{3r}r(1+r)+m_{3r}^{2}r^{2}(1+r)^{2}} \right. \Bigg] \nonumber.
\end{eqnarray}
For example, if $m_{3}$ is constant so that $m_{3r}=0$, the eigenvalues reduce to, 
\begin{equation*}
\frac{3+2\beta}{1+2\beta}, \; \frac{6+8\beta}{1+2\beta}, \; \frac{6m_{3}+4\beta(-1+m_{3})}{m_{3}(1+2\beta)}.
\end{equation*}
 In the range $-3/4 < \beta < -1/2$ the first and second eigenvalues are negative, in this  constant $m_{3}$ case. To get stable acceleration, one must find corresponding values of $m_{3}$ so as to get the third eigenvalue negative or zero as well. 

		 	
\vspace{0.25cm}
\noindent\underline {Point $\mathbbm{P}4$:} 
 
\noindent The eigenvalues for $\mathbbm{P}4$ are,
\begin{equation}
	-3, \; -2-\frac{3}{3-8\beta^{2}}, \; \frac{16\beta^{2}(1+m_{4})}{m_{4}(3-8\beta^{2})}, 
	\label{EV4}
\end{equation}
where $m_{4}=m(r=0)$. Accelerative expansion occurs for $\beta < -\sqrt{3/8}, \; \beta > \sqrt{3/8}$, or $|\beta| < \sqrt{3}/4$. The first eigenvalue is negative, and the second one is negative or equal to zero for $\beta \leq -3/4, \; \beta \geq 3/4$, or $|\beta| < \sqrt{3/8}$. In the overlapping regions, \textit{i.e.} when $\beta \leq -3/4, \; \beta \geq 3/4, \; |\beta| < \sqrt{3}/4$ we have almost all conditions for stable acceleration satisfied. We can then find corresponding values of $m_{4}$ so that the third eigenvalue also becomes negative or zero. For $|\beta| < \sqrt{3}/4$ and $-1 \leq m(r=0) < 0$, $\mathbbm{P}4$ always represents a stable accelerated era with $w_{eff}>-1$, with no condition on $m_{4r}$. The point $\mathbbm{P}4$ is therefore of interest and we discuss it further in sections \ref{Fourd} and \ref{Foure}. 

			
\vspace{0.25cm}
\noindent\underline{Point $\mathbbm{P}5$:} 

\noindent The general expression for the eigenvalues of $\mathbbm{P}5$ is non-trivial. 
The eigenvalues of $\mathbbm{P}5$ in the limit of a matter dominated era (with $|m_{5}|<<1$) are approximately,
\begin{equation*}
3(1+m_{5r}), \; -\frac{3}{4} \pm \sqrt{-\frac{4\beta^{2}}{m_{5}}}.
\end{equation*}

Models with $m_{5}<0$ are not acceptable since the	eigenvalues diverge as $m_{5} \rightarrow 0^{-}$, so the system would not remain at $\mathbbm{P}5$ for sufficient time to give rise to a viable CDM dominated era \cite{Amendola:2006we}. In order to get a valid saddle MDE as $m_{5} \rightarrow 0^{+}$ we need,
\begin{equation*}
m(r \leq -1) > 0, \; m_{r}(r \leq -1) > -1, \; m(r = -1) = 0.
\end{equation*}

We find that $\mathbbm{P}5$ is never stable in either of the two regions that give accelerated expansion ($m_5<-1$ and $m_5>1/2$), for values of $\beta$ approximately in the range $|\beta| \leq 0.86$. For $\beta$ outside of this range $\mathbbm{P}5$ can give rise to accelerated expansion, however these values of $\beta$ are generally disfavored by observations \cite{Amendola:1999er}. Hence $\mathbbm{P}5$ best describes a MDE. 

			
\vspace{0.25cm}
\noindent\underline{Point $\mathbbm{P}6$:} 

\noindent The eigenvalues for $\mathbbm{P}6$ are,
\begin{eqnarray}
&& \frac{-4\beta^{2} + 8\beta^{2}m_{6} + m_{6}^{2}(9-4\beta^{2})}{m_{6}[-4\beta^{2} + m_{6}(-3+4\beta^{2})]}, \; \frac{-8\beta^{2} +	12\beta^{2}m_{6} + m_{6}^{2}(9-4\beta^{2})}{m_{6}[-4\beta^{2} + m_{6}(-3+4\beta^{2})]},\nonumber \\
&& \frac{8\beta^{2}(-1+m_{6}^{2})(1+m_{6r})}{m_{6}[-4\beta^{2} + m_{6}(-3+4\beta^{2})]}.
\label{EV6}
\end{eqnarray}
We note that  there is a symmetry with respect to positive and negative values of $\beta$ since $w_{eff}$ and the eigenvalues are all functions of $\beta^{2}$.

In order to characterize and study the properties of $\mathbbm{P}6$ in more detail, we limit our analysis to $\beta$ in the range $|\beta| \leq 1/2$, in which we find that it is accelerated and stable in five distinct ranges. This choice of $\beta$ is well-motivated, the value of $\beta$ is constrained by CMB observations with $|\beta| \leq 0.1$ in \cite{Amendola:1999er} and encompasses $f(R)$ theories with $\beta=1/2$. Hence $|\beta| \leq 1/2$ represents a broad and interesting range for viable scalar-tensor theories. 

There are five regions of stable acceleration in $\mathbbm{P}6$, labelled (A)-(E), which depend on the magnitude of $m$ at the fixed point, $m_6$, and the sign of the derivative $dm/dr$ at the fixed point, written  $m_{6r}$.
{\renewcommand{\labelenumi}{[\Roman{enumi}]}
\renewcommand{\labelenumii}{(\Alph{enumii})}
\begin{enumerate}
\item $m_{6r}>-1$ 

\begin{enumerate}
\item $m_{6} \leq$ min $[-1,2|\beta|/(2|\beta| - \sqrt{3})]$: \ \ $w_{eff} > -1$,
\item $4\beta^{2}/(4\beta^{2}-3) < m_{6} < 0$: \ \ \ \  \ \ \ \ \ \ \ \  \ $w_{eff}\ll -1$,
\item $m_{6} \geq 1$: \ \ \ \ \ \ \ \ \ \ \ \ \ \ \ \ \ \ \ \ \ \ \ \ \ \ \ \ \ \ \ \ \ \ \ $w_{eff} \lesssim -1$,
\end{enumerate}	

\item $m_{6r}<-1$ 

\begin{enumerate}[resume]
\item $2|\beta|/(2|\beta| + \sqrt{3}) < m_{6} \leq 1$: $w_{eff} > -1$
\item $-1 \leq m_{6} < 2|\beta|/(2|\beta| - \sqrt{3})$: This point exists only for $\beta$ in the range $|\beta| < \sqrt{3}/4$ for which $w_{eff}>-1$.
\end{enumerate}	
\end{enumerate}
}

Region (E) does not exist for $f(R)$ theories, so as with $\mathbbm{P}4$, general values of $\beta$ therefore open up avenues for new accelerative critical points. We discuss this region further in sections \ref{Fourd} and \ref{Foure}.


\vspace{0.25cm}
\noindent\underline{Point $\mathbbm{P}7$:} 

\noindent The eigenvalues for $\mathbbm{P}7$ are,
\begin{eqnarray}
	\fl \frac{3-2\beta}{1-2\beta}, \ \ \frac{2}{m_{7}^{2}(1-2\beta)} \Bigg[ \beta m_{7} - \beta m_{7r}r(1+r) - 3m_{7}^{2}(-1+\beta) \nonumber \\
	\fl \ \ \pm \beta \sqrt{m_{7}^{4}+2m_{7}^{3}+m_{7}^{2}\left\{1+2m_{7r}r(-1+r)\right\}-2m_{7} m_{7r}r(1+r)+m_{7r}^{2}r^{2}(1+r)^{2}} \Bigg]. \nonumber
\end{eqnarray}
For example, if $m_{7}$ is constant so that $m_{7r}=0$, the eigenvalues reduce to, 
\begin{equation*}
\frac{3-2\beta}{1-2\beta}, \; \frac{6-8\beta}{1-2\beta}, \; \frac{6m_{7}+4\beta(1-m_{7})}{m_{7}(1-2\beta)}
\end{equation*}
Stable acceleration is able to occur for the range $1/2 < \beta < 3/4$, where the first and second eigenvalues are negative (in the constant $m_{7}$ case), and a suitable value of $m_{7}$ can make the third eigenvalue to also  be negative or zero.

\subsection{Classification of $F(\Phi)R$ scalar-tensor theories}
\label{Fourd}

We classify general $F(\Phi)R$ scalar-tensor theories, with values of $\beta$ in the range $|\beta| \leq 1/2$, on the basis of trajectories of the $m(r)$ line on the $(r,m)$ plane since such a trajectory completely specifies a theory's dynamical evolution as the universe expands. These trajectories depend on the functional form of $F(\Phi)$, or from an alternative perspective on the value of $\beta$. We therefore consider conditions on viable cosmologies as conditions on these trajectories.

\begin{itemize}
\item 
We assume that the function $f$ is a $C^{\infty}$ function (\textit{i.e.} it is differentiable for all degrees of differentiation). Also $f$ and all of its derivatives are non-singular and single valued. For the conformal transformation and scalar field redefinition to be valid we require that $F(\Phi) > 0$. This implies that $f(\Phi)$ is monotonically increasing and single valued for all $\Phi$. As such $r(\Phi)$ and $m(\Phi)$ are also single valued. Therefore $m(r)$ is single valued and we can consider evolutionary trajectories in terms of the function $m(r)$.

\item As mentioned earlier, a viable cosmology requires an extended matter dominated era followed by late time acceleration. The matter dominated era therefore needs to be a saddle point. From our detailed analysis in Sec. \ref{Fourc} we find that only the point $\mathbbm{P}5$ with $m_{5}(r=-1) \rightarrow 0^{+}$ can be used as a standard saddle matter dominated point with $\Omega_{m}=1$, $w_{eff}=0$ and $a \propto t^{2/3}$. This point is denoted as $\mathbbm{P}5^{(0)}$ . 

\item The accelerated expansion needs to be a stable or saddle point. Valid acceleration points (for $|\beta| \leq 1/2$) are $\mathbbm{P}1$, $\mathbbm{P}4$, and $\mathbbm{P}6$. Transitions from $\mathbbm{P}5^{(0)}$ to $\mathbbm{P}1$ and regions (A)-(D) of $\mathbbm{P}6$ are qualitatively similar to those in $f(R)$ theories \cite{Amendola:2006we}. The point $\mathbbm{P}4$ and the region (E) of point $\mathbbm{P}6$, however, are not present in $f(R)$ models, we therefore discuss their properties in detail below. Interestingly, both of these exist as stable accelerated epochs only for $|\beta| < \sqrt{3}/4$ and have very similar properties. 

\item The points $\mathbbm{P}5$ and $\mathbbm{P}6$ lie on the critical line, $m=-r-1$. Valid $m(r)$ trajectories should not enter a critical point from a forbidden direction. For the point $\mathbbm{P}5^{(0)}$ to exist we require $m_{5r} > -1$. We can therefore mark \textit{forbidden direction regions} on the critical line around the point $\mathbbm{P}5^{(0)}$.  Forbidden direction regions also exist around $\mathbbm{P}6$, with transitions to the regions (A), (B) and (C) of point $\mathbbm{P}6$ requiring that $m_{6r}>-1$, and $m_{6r}<-1$ for transitions to regions (D) and (E). The point $\mathbbm{P}4$, by contrast, has no such forbidden direction regions around it. 

\end{itemize}
The characteristics of the 5 general classes of scalar-tensor theory trajectories that give stable acceleration are summarized below. 

\begin{enumerate}

\item[\textbf{I}:] The $m(r)$ curve does not connect the standard matter dominated point to the accelerated attractor solutions. Models of this class either bypass the matter era and directly fall onto an accelerated attractor or go through a $\phi$MDE. For general values of $\beta$, however, $w_{eff}$ for the point $\mathbbm{P}2$ can be arbitrarily close to zero and give a standard MDE instead of a $\phi$MDE. 

\item[\textbf{II}:]  The $m(r)$ curve connects the standard matter era $\mathbbm{P}5^{(0)}$ to $\mathbbm{P}1$ or asymptotically ($r \rightarrow  \pm \infty$) to $\mathbbm{P}6$, giving a stable accelerated de-Sitter expansion. 

\item[\textbf{III}:]  The $m(r)$ curve intersects the critical line in the region (B) of point $\mathbbm{P}6$. These models have very short matter dominated phases which would not allow sufficient time for structure formation. The resulting acceleration after matter domination has $w_{eff}\ll -1$. 

\item[\textbf{IV}:]  The $m(r)$ curve connects the standard matter era $\mathbbm{P}5^{(0)}$ to the region (D) of $\mathbbm{P}6$ which has acceleration with $w_{eff}>-1$.

\item[\textbf{V}:]  Models of this class connect the standard MDE $\mathbbm{P}5^{(0)}$ to the point $\mathbbm{P}4$ in $-1 \leq m(r=0) < 0$, or to region (E) of $\mathbbm{P}6$. These models exist only for $|\beta| < \sqrt{3}/4$, hence $\mathbbm{P}4$ and region (E) of $\mathbbm{P}6$ are not accelerated for $f(R)$ theories. The final acceleration has $w_{eff}>-1$.
 
\end{enumerate}

Classes I-IV are qualitatively similar to those in $f(R)$ theories. Class V is not open to $f(R)$ theories and we discuss this class here in more detail. 

Class V covers the transition between points $\mathbbm{P}5^{(0)}$ and $\mathbbm{P}4$ or region (E) of $\mathbbm{P}6$. As shown in Sec. \ref{Fourc} the point $\mathbbm{P}4$ and region (E) of $\mathbbm{P}6$ represent accelerated epochs with $w_{eff}>-1$ for $|\beta| < \sqrt{3}/4$. $\mathbbm{P}4$ is stable for $-1 \leq m(r=0) < 0$ irrespective of the value of $m_{4r}$, and region (E) of $\mathbbm{P}6$ (which has $-1 \leq m_{6} < 2|\beta|/(2|\beta| - \sqrt{3})$) is stable for $m_{6r}<-1$. If we can find a function $f(\Phi)$ which takes us from $\mathbbm{P}5^{(0)}$ to either of these points then this would represent a valid trajectory for cosmic evolution. 

\begin{figure}[t]
\centering{
\hbox{\includegraphics[width=6.7cm]{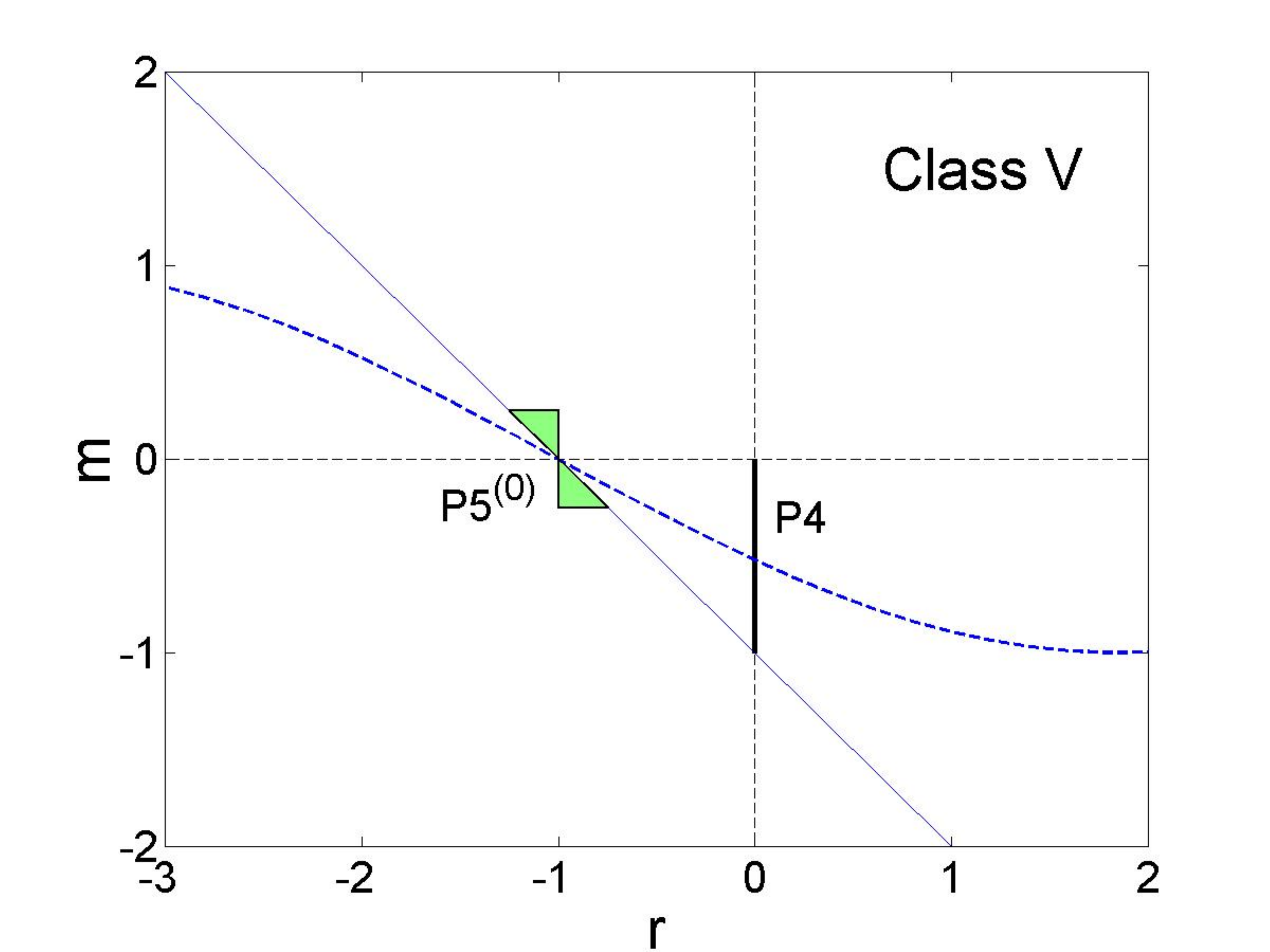}\includegraphics[width=6.7cm]{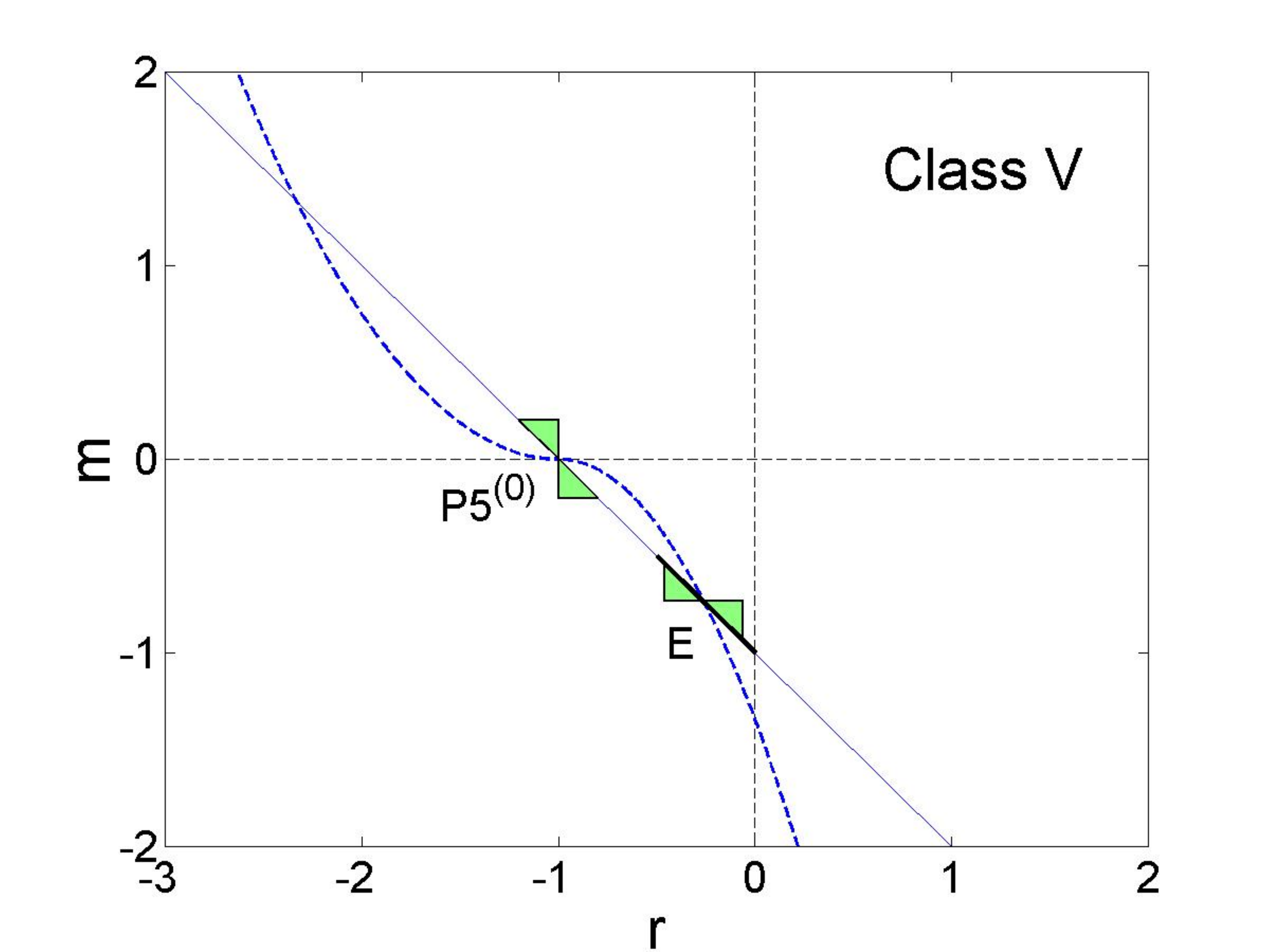}}
\caption{Class V models of scalar-tensor theories on the $(r,m)$ plane. The solid line with a slope of -1 is the critical line $m=-r-1$ and the dashed curves are possible trajectories for a Class V model. The point $\mathbbm{P}5^{(0)}$ $(-1,0)$ is the starting saddle matter dominated point, with the triangles around it representing the forbidden direction regions. The $m(r)$ curve for a Class V model intersects either the $m$-axis between 0 and -1 corresponding to the point $\mathbbm{P}4$, or the critical line in region (E) of point $\mathbbm{P}6$, to get a stable accelerated epoch.}
\label{graph2}
}
\end{figure}

In Figure \ref{graph2} we show possible trajectories on the $(r,m)$ plane for a Class V model. Consider the case when $\mathbbm{P}5^{(0)}$ is connected to $\mathbbm{P}4$ by a straight line. We require the slope of this line to be between 0 and -1 so that it can intersect the $m$-axis between 0 and -1. The resulting equation for the $m(r)$ curve that we are looking for is,
\begin{equation}
m = -\frac{1}{n}(1+r)\label{mconstraint}
\end{equation}
where $n>1$. Combining (\ref{mconstraint}) with (\ref{Fdef}), (\ref{mdef}) and (\ref{rdef}),  the resulting second-order differential equation in $f(\Phi)$ gives solutions for Class V models of the form,
\begin{equation}
f(\Phi) = a(\Phi^{\frac{n-1}{n}} + b)^{\frac{n}{n-1}}
\end{equation} 
where $a$ and $b$ are constants, and $a>0$. This gives, 
\begin{equation}
r = -\frac{\Phi}{\Phi + b\Phi^{1/n}}
\end{equation}
In order to have the correct sequence of matter domination and acceleration we need $r$ to be increasing with time, \textit{i.e.}  $\textrm{d}r/\textrm{d}a > 0$. Using $\dot{\Phi}<0$ and $\dot{a}>0$ this corresponds to $b>0$ (here dots denote derivatives with physical time in the JF). In $f(R)$ theories, $\dot{\Phi}<0$ represents a curvature, $R$, that decreases with time consistent with observations. 

The above form of $f(\Phi)$ can be used to calculate the mass of the scalar field $\Phi$ in these scalar-tensor theories \cite{Chiba:2003ir}, 
\begin{equation}
m_{\phi}^{2} = \tilde{V}_{,\phi\phi} |_{\phi=0} \; = \; \frac{4\beta^{2}}{3} \left( \frac{d\Phi}{dF} + \Phi - 4f(\Phi) \right) 
\end{equation}

For large values of $\Phi$, $b>0$ corresponds to $m_{\phi}^{2} < 0$. Thus this simple $f(\Phi)$ leads to scalar fields with imaginary mass. If we have some general curve joining $\mathbbm{P}5^{(0)}$ and $\mathbbm{P}4$ we can break it up into an infinite number of straight lines with at least some being of the form of the above $f(\Phi)$ (with slope between -1 and 0) and again argue that it would lead to an imaginary mass for the scalar field. A transition from $\mathbbm{P}5^{(0)}$ to region (E) of $\mathbbm{P}6$ will similarly need an imaginary mass  scalar field because this transition also goes in the direction of increasing $r$ with a negative value of $m$. Hence Class V have a matter dominated era followed by an accelerated epoch with $w_{eff}>-1$, but relate to theories with imaginary scalar field masses in the Einstein frame.  In the specfic case of $f(R)$ models this translates to imaginary mass scalar fields arising when $f_{RR}<0$ - see Refs. \cite{Song:2006ej,Sawicki:2007tf} for details.  

\subsection{Criteria for viable models with saddle acceleration}
\label{Foure}

Till now we have only studied viable cosmologies which go from a saddle matter dominated era to a stable accelerated universe. Even though we know that dark energy is the dominant component today we can not be sure of what the future holds. Therefore it may be possible that acceleration is not a permanent feature of our universe \cite{Barrow:2000nc}. 

We first consider acceleration that is stable on the subspace $\tilde{z}=r\tilde{y}$, on which saddle point CDM attractors are stable, where $r$ (and hence $m$) is constant. The evolution equations written earlier reduce to,
\begin{eqnarray}
\tilde{x}' & = & \beta[1-\tilde{x}^{2}-(1+r)\tilde{y}] + 2\beta(2+r)\tilde{y} - \frac{3}{2}\tilde{x}[1-\tilde{x}^{2}+(1+r)\tilde{y}] \\
\tilde{y}' & = & - 4\beta r\tilde{x}\tilde{y} \left( \frac{1}{m} \right) - 8\beta \tilde{x}\tilde{y} + 3\tilde{y}[1+\tilde{x}^{2}-(1+r)\tilde{y}]
\end{eqnarray}
\\
where $r=-(1+m)$. The solutions of these evolution equations are the points $\mathbbm{P}2$, $\mathbbm{P}3$, $\mathbbm{P}5$, $\mathbbm{P}6$, and $\mathbbm{P}7$. Since $m$ is now constant, the points $\mathbbm{P}1$ and $\mathbbm{P}4$ are just special cases of $\mathbbm{P}6$ with $m=\pm 1$. 

For $|\beta| \leq 1/2$ we see that the only points that can be accelerated are $\mathbbm{P}5$ and $\mathbbm{P}6$. It can be shown that $\mathbbm{P}5$ is not stable (even on the subspace $\tilde{z}=r\tilde{y}$) in the regions that it is accelerated. The point $\mathbbm{P}6$ $\left( \frac{2\beta(-1+m)}{3m}, \frac{4\beta^{2} - 8\beta^{2}m + m^{2}(-9+4\beta^{2})}{9m^{3}} \right)$ has $\Omega_{m} = 0$, an effective equation of state, 
\begin{equation}
	\frac{-8\beta^{2}+20\beta^{2}m+m^{2}(9-12\beta^{2})}{3m[-4\beta^{2}+m(-3+4\beta^{2})]},
\end{equation}
and its eigenvalues are given by the first two in (\ref{EV6}),
\begin{eqnarray}
	\frac{-4\beta^{2} + 8\beta^{2}m_{6} + m_{6}^{2}(9-4\beta^{2})}{m_{6}[-4\beta^{2} + m_{6}(-3+4\beta^{2})]}, \; \frac{-8\beta^{2} +	12\beta^{2}m_{6} + m_{6}^{2}(9-4\beta^{2})}{m_{6}[-4\beta^{2} + m_{6}(-3+4\beta^{2})]}
\end{eqnarray}
$\mathbbm{P}6$ is accelerated ($w_{eff}<-1/3$) and stable on the subspace $\tilde{z}=r\tilde{y}$ in the following three regions:

\renewcommand{\labelenumi}{(\Alph{enumi})}
\begin{enumerate}
\item $m_{6} < 2|\beta|/(2|\beta| - \sqrt{3})$: $w_{eff} > -1$
\item $4\beta^{2}/(4\beta^{2}-3) < m_{6} < 0$: $w_{eff}\ll -1$
\item $m_{6} > 2|\beta|/(2|\beta| + \sqrt{3})$: $w_{eff} > -1$ for $m_{6} < 1$, and $w_{eff}\lesssim-1$ for $m_{6} \geq 1$
\end{enumerate}

\begin{figure}[t]
\centering{
\hbox{\includegraphics[width=6.7cm]{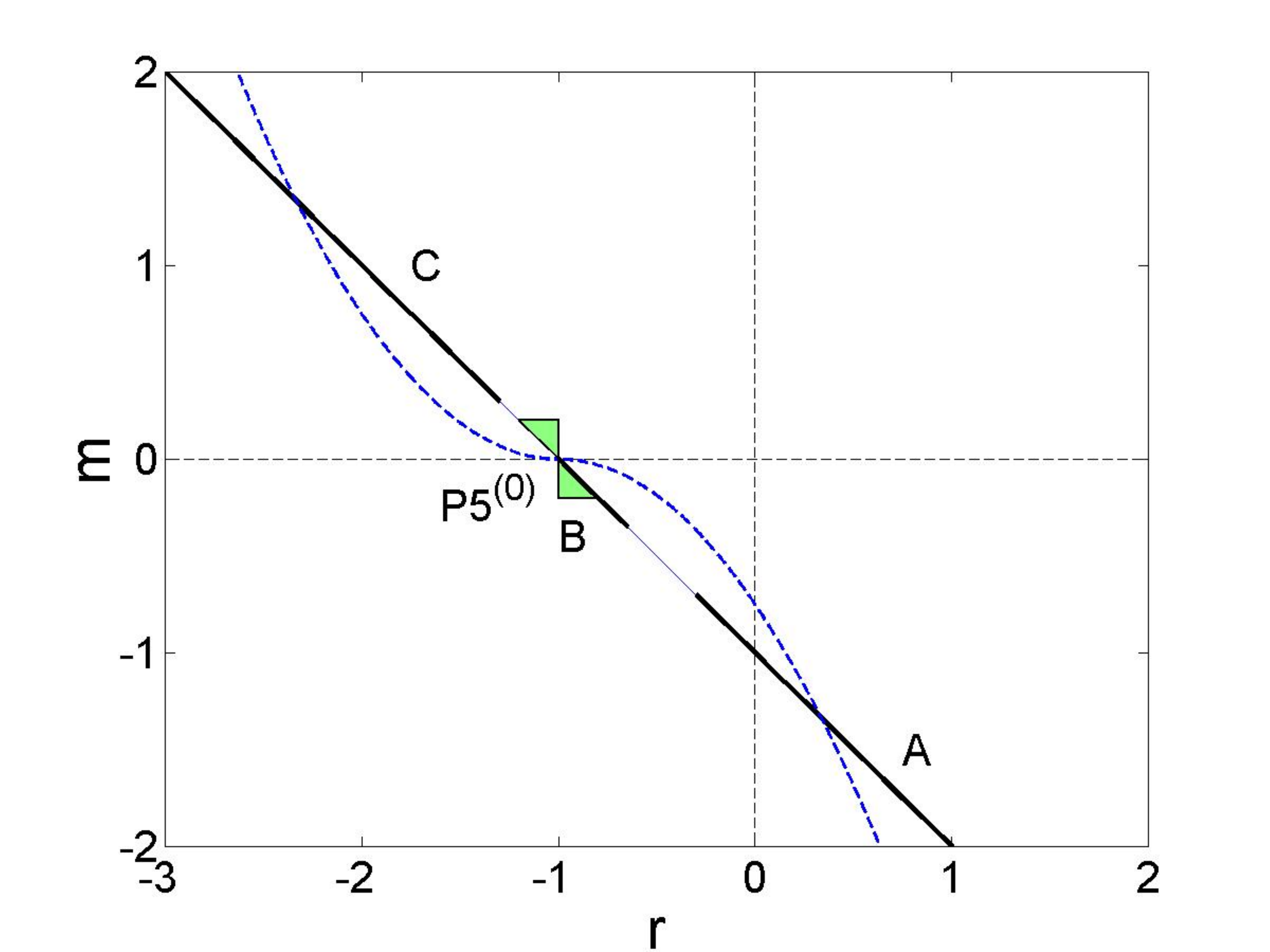}\includegraphics[width=6.7cm]{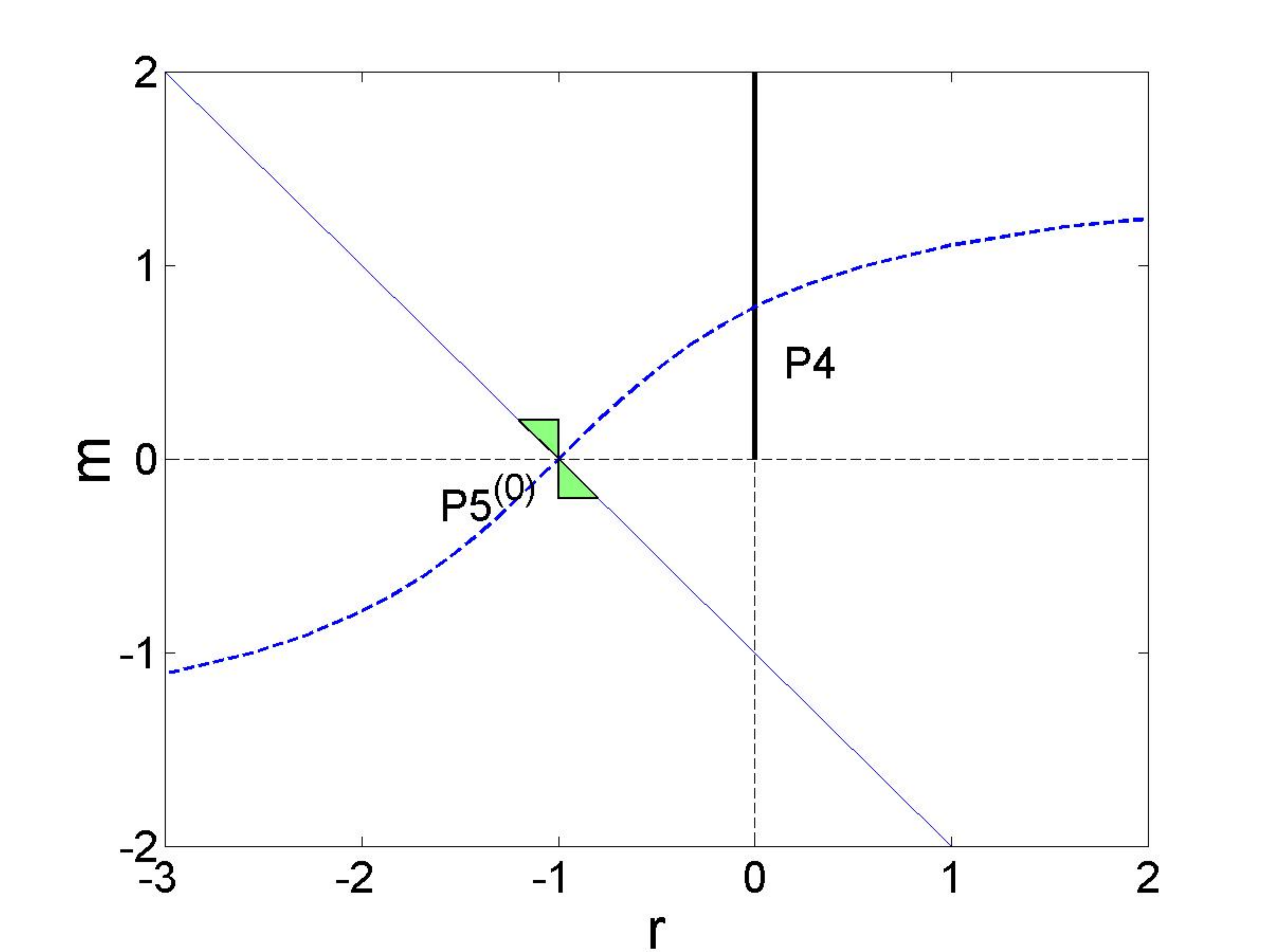}}
\caption{Scalar-tensor theories which lead to a saddle accelerated expansion of the universe. The solid line with a slope of -1 is the critical line $m=-r-1$ and the dashed curves are possible trajectories for a scalar-tensor theory. The point $\mathbbm{P}5^{(0)}$ $(-1,0)$ is the starting saddle matter dominated point, with the triangles around it representing the forbidden direction regions. In the first graph the $m(r)$ curve connects the matter dominated point to one of the three accelerated regions (A, B or C) of the point $\mathbbm{P}6$. The accelerated expansion is stable on the subspace $\tilde{z}=r\tilde{y}$. In the second graph the $m(r)$ curve connects the matter dominated point to the point $\mathbbm{P}4$ on $m(r=0)>0$ which gives a period of, potentially transient, acceleration.} \label{graph3}
}
\end{figure}

Notice that there is no condition on $m_{6r}$, therefore there is no forbidden direction region around $\mathbbm{P}6$. The first graph in Figure~\ref{graph3} shows a possible trajectory of the $m(r)$ curve, connecting $\mathbbm{P}5^{(0)}$ to either of the regions (A) and (C). We can also connect to the region (B) of $\mathbbm{P}6$. It is likely that all scalar-tensor theories which connect to the regions (A) and (B) are tachyonic because of reasons discussed earlier (since they have negative values of $m$ and are moving in the direction of increasing $r$). This only leaves region (C) to be a viable cosmology. Theories which connect $\mathbbm{P}5^{(0)}$ to region (C) of $\mathbbm{P}6$ are therefore possible models for the evolution of the universe from a saddle matter dominated era to a saddle accelerated expansion. 

On a general subspace (away from $\tilde{z} = r \tilde{y}$) the point $\mathbbm{P}4$ admits acceleration for $|\beta| < \sqrt{3}/4$, not compatible with $f(R)$ theories, that gives rise to a second saddle acceleration attractor. If we consider positive values of $m(r=0)$ then we see that the third eigenvalue in (\ref{EV4}) is positive but the first two are still negative. Thus $\mathbbm{P}4$ with $m(r=0)>0$ represents saddle acceleration on such a subspace, for $|\beta| < \sqrt{3}/4$. The second graph in Figure~\ref{graph3} shows a possible trajectory from the matter dominated point $\mathbbm{P}5^{(0)}$ to the saddle acceleration point $\mathbbm{P}4$.

\section{Conclusions}
\label{Five}

In this paper we have established the dynamical attractor behavior in scalar-tensor theories of dark energy, presenting a complete, consistent picture of evolution in both the Einstein and Jordan frames. We discuss critical points for the dynamical evolution and show how in the two frames the stability matrices for these points are related by a similarity transformation.

We carry out a general analysis for values of the coupling $\beta$ in $|\beta| \leq 1/2$, and find that there are five classes of evolutionary behavior, of which four classes are qualitatively similar to those for $f(R)$ gravity. The fifth class only exists for values of $\beta$ in the range $|\beta| < \sqrt{3}/4$, \textit{i.e.} it is not present for $f(R$) gravity. This class of models has a standard matter dominated phase followed by acceleration with $w_{eff}>-1$, but for stable acceleration seems to always have tachyonic scalar fields. 

We then relax the condition of a stable accelerated universe and study transitions from the allowed MDE to a saddle accelerated universe, again for $|\beta| \leq 1/2$. We find that on the subspace $\tilde{z}=r\tilde{y}$ ($r$ = constant), which contains the suitable MDE, the point $\mathbbm{P}6$ is accelerated and gives evolution without tachyonic scalar fields if $m_{6} > 2|\beta|/(2|\beta| + \sqrt{3})$. This class of behavior is qualitatively true also for $f(R)$ theories ($\beta=1/2$). In addition there is a possibility of going from the MDE to the saddle non-phantom acceleration point $\mathbbm{P}4$, only open to $|\beta|<\sqrt{3}/4$. Therefore if we allow the current accelerated expansion of the universe to be temporary then there are a much broader variety of $f(R)$ and $F(\Phi)R$ theories that are dynamically valid.

Our results open up interesting questions about broader dark energy theories. In particular, in this paper we have used a particular form of the potential in the EF given by Eq. (\ref{potchoice}) appropriate for a wide class of theories, including $f(R)$ theories, in which the Jordan frame action has no scalar potential. It will be interesting to see the implications of using a more general form of the potential. 

\section*{Acknowledgements}

We thank Istvan Laszlo for valuable discussions in the course of this work. The work of NA and RB is supported by the National Science Foundation under grants AST-0607018 and PHY-0555216.

\section*{References}
\bibliographystyle{iopart-num}
\bibliography{paper}

\end{document}